# Challenges and Practices in Aligning Requirements with Verification and Validation: A Case Study of Six Companies

Elizabeth Bjarnason[1], Per Runeson[1], Markus Borg[1], Michael Unterkalmsteiner[2], Emelie Engström[1], Björn Regnell[1], Giedre Sabaliauskaite[5], Annabella Loconsole[3], Tony Gorschek[2], Robert Feldt[2,4]

[1]*Lund University, Sweden*
[2]*Blekinge Institute of Technology, Sweden*
[3]*Malmö University, Sweden*
[4]*Chalmers University of Technology, Sweden*
[5]*Singapore University of Technology and Design, Singapore*

**Abstract:** Weak alignment of requirements engineering (RE) with verification and validation (VV) may lead to problems in delivering the required products in time with the right quality. For example, weak communication of requirements changes to testers may result in lack of verification of new requirements and incorrect verification of old invalid requirements, leading to software quality problems, wasted effort and delays. However, despite the serious implications of weak alignment research and practice both tend to focus on one or the other of RE or VV rather than on the alignment of the two. We have performed a multi-unit case study to gain insight into issues around aligning RE and VV by interviewing 30 practitioners from 6 software developing companies, involving 10 researchers in a flexible research process for case studies. The results describe current industry challenges and practices in aligning RE with VV, ranging from quality of the individual RE and VV activities, through tracing and tools, to change control and sharing a common understanding at strategy, goal and design level. The study identified that human aspects are central, i.e. cooperation and communication, and that requirements engineering practices are a critical basis for alignment. Further, the size of an organisation and its motivation for applying alignment practices, e.g. external enforcement of traceability, are variation factors that play a key role in achieving alignment. Our results provide a strategic roadmap for practitioners improvement work to address alignment challenges. Furthermore, the study provides a foundation for continued research to improve the alignment of RE with VV.

# 1 Introduction

Requirements engineering (RE) and verification and validation (VV) both aim to support development of products that will meet customers' expectations regarding functionality and quality. However, to achieve this RE and VV need to be aligned and their *'activities or systems organised so that they match or fit well together'* (MacMillan Dictionary's definition of 'align'). When aligned within a project or an organisation, RE and VV work together like two bookends that support a row of books by buttressing them from either end. RE and VV, when aligned, can effectively support the development activities between the initial definition of requirements and acceptance testing of the final product (Damian 2006).

Weak coordination of requirements with development and testing tasks can lead to inefficient development, delays and problems with the functionality and the quality of the produced software, especially for large-scale development (Kraut 1995). For example, if requirements changes are agreed without involving testers and without updating the requirements specification, the changed functionality is either not verified or incorrectly verified. This weak alignment of RE and work that is divided and distributed among engineers within a company or project poses a risk of producing a product that does not satisfy business and/or client expectations (Gorschek 2007). In particular, weak alignment between RE and VV may lead to a number of problems that affect the later project phases such as non-verifiable requirements, lower product quality, additional cost and effort required for removing defects (Sabaliauskaite 2010). Furthermore, Jones et al. (2009) identified three other alignment related problems found to affect independent testing teams, namely uncertain test coverage, not knowing whether changed software behaviour is intended, and lack of established communication channels to deal with issues and questions.

There is a large body of knowledge for the separate areas of RE and VV, some of which touches on the connection to the other field. However, few studies have focused specifically on the alignment between the two areas (Barmi 2011) though there are some exceptions. Kukkanen et al. reported on lessons learnt in concurrently improving the requirements and the testing processes based on a case study (Kukkanen 2009). Another related study was performed by Uusitalo et al. who identified a set of practices used in industry for linking requirements and testing (Uusitalo 2008). Furthermore, RE alignment in the context of outsourced development has been pointed out as a focus area for future RE research by Cheng and Attlee (Cheng 2007).

When considering alignment, traceability has often been a focal point (Watkins 1994, Barmi 2011, Paci 2012). However, REVV alignment also covers the coordination between roles and activities of RE and VV. Traceability mainly focuses on the structuring and organisation of different related artefacts. Connecting (or tracing) requirements with the test cases that verify them support engineers in ensuring requirements coverage, performing impact analysis for requirements changes etc. In addition to tracing, alignment also covers the interaction between roles throughout different project phases; from agreeing on high-level business and testing strategies to defining and deploying detailed requirements and test cases.

Our case study investigates the challenges of RE and VV (REVV) alignment, and identifies methods and practices used, or suggested for use, by industry to address these issues. The results reported in this paper are based on semi-structured interviews of 90 minutes each with 30 practitioners from six different software companies, comprising a wide range of people with experience from different roles relating to RE and VV. This paper extends on preliminary results of identifying the challenges faced by one of the companies included in our study (Sabaliauskaite 2010). In this paper, we report on the practices and challenges of all the included companies based on a full analysis of all the interview data. In addition, the results are herein categorised to support practitioners in defining a strategy for identifying suitable practices for addressing challenges experienced in their own organisations.

The rest of this paper is organised as follows: Section 2 presents related work. The design of the case study is described in Section 3, while the results can be found in Section 4. In Section 5 the results are discussed and, finally the paper is concluded in Section 6.

# 2 Related Work

The software engineering fields RE and VV have mainly been explored with a focus on one or the other of the two fields (Barmi 2011), though there are some studies investigating the alignment between the two. Through a systematic mapping study into alignment of requirements specification and testing, Barmi et al. found that most studies in the area were on model-based testing including a range of variants of formal methods for describing requirements with models or languages from which test case are then generated. Barmi et al. also identified traceability and empirical studies into alignment challenges and practices as main areas of research. Only 3 empirical studies into REVV alignment were found. Of these, 2 originate from the same research group and the third one is the initial results of the study reported in this paper. Barmi et al. draw the conclusions that though the areas of model-based engineering and traceability are well understood, practical solutions including evaluations of the research are needed. In the following sections previous work in the field is described and related to this study at a high level. Our findings in relation to previous work are discussed in more depth in Section 5.

**The impact of RE on the software development process** as a whole (including testing) has been studied by Damian et al. (2005) who found that improved RE and involving more roles in the RE activities had positive effects on testing. In particular, the improved change control process was found to 'bring together not only the functional organisation through horizontal alignment (designers, developers, testers and documenters), but also vertical alignment of organisational responsibility (engineers, teams leads, technical managers and executive management)' (Damian 2005). Furthermore, in another study Damian and Chisan (2006) found that rich interactions between RE and testing can lead to pay-offs in improved test coverage and risk management, and in reduced requirements creep, overscoping and waste, resulting in increased productivity and product quality. Gorschek and Davis (2007) have proposed a taxonomy for assessing the impact of RE on, not just project, but also on product, company and society level; to judge RE not just by the quality of the system requirements specification, but also by its wider impact.

**Jointly improving the RE and testing processes** was investigated by Kukkanen et al. (2009) through a case study on development performed partly in the safety-critical domain with the dual aim of improving customer satisfaction and product quality. They report that integrating requirements and testing processes, including clearly defining RE and testing roles for the integrated process, improves alignment by connecting processes and people from requirements and testing, as well as, applying good practices that support this connection . Furthermore, they report that the most important aspect in achieving alignment is to ensure that 'the right information is communicated to the right persons' (Kukkanen 2009, p. 484). Successful collaboration between requirements and test can be ensured by assigning and connecting roles from both requirements and test as responsible for ensuring that reviews are conducted. Among the practices implemented to support requirements and test alignment were the use of metrics, traceability with tool support, change management process and reviews of requirements, test cases and traces between them (Kukkanen 2009). The risk of overlapping roles and activities between requirements and test, and gaps in the processes was found to be reduced by concurrently improving both processes (Kukkanen 2009). These findings correlate very well with the practices identified through our study.

**Alignment practices** that improve the link between requirements and test are reported by Uusitalo et al. (2008) based on six interviews, mainly with test roles, from the same number of companies. Their results include a number of practices that increase the communication and interaction between requirements and testing roles, namely early tester participation, traceability policies, consider feature requests from testers, and linking test and requirements people. In addition, four of the companies applied traceability between requirements and test cases, while admitting that traces were rarely maintained and were thus incomplete (Uusitalo 2008). Linking people or artefacts were seen as equally important by the interviewees who were unwilling to select one over the other. Most of the practices reported by Uusitalo et al. were also identified in our study with the exception of the specific practice of linking testers to requirements owners and the practice of including internal testing requirements in the project scope.

**The concept of traceability** has been discussed, and researched since the very beginning of software engineering, i.e. since the 1960s (Randell 1969). Traceability between requirements and other development artefacts can support impact analysis (Gotel 1994, Watkins 1994, Ramesh 1997, Damian 2005, Uusitalo 2008, Kukkanen 2009), lower testing and maintenance costs (Watkins 1994, Kukkanen 2009), and increased test coverage (Watkins 1994, Uusitalo 2008) and thereby quality in the final products (Watkins 1994, Ramesh 1997). Tracing is also important to software verification due to being an (acknowledged) important aspect in high quality development (Watkins 1994, Ramesh 1997). The challenges connected to traceability have been empirically investigated and reported over the years. The found challenges include volatility of the traced artefacts, informal processes with lack of clear responsibilities for tracing, communication gaps, insufficient time and resources for maintaining traces in combination with the practice being seen as non-cost efficient, and a lack of training (Cleland-Huang 2003). Several methods for supporting automatic or semi-automatic recovery of traces have been proposed as a way to address the cost of establishing and maintaining traces, e.g. De Lucia 2007, Hayes 2007, Lormans 2008. An alternative approach is proposed by Post et al. (2009) where the number of traces between requirements and test are reduced by linking test cases to user scenarios abstracted from the formal requirements, thus tracing at a higher abstraction level. When evaluating this approach, errors were found both in the formal requirements and in the developed product (Post 2009). However, though the evaluation was performed in an industrial setting the set of 50 requirements was very small. In conclusion, traceability in full-scale industrial projects remains an elusive and costly practice to realise (Gotel 1994, Watkins 1994, Jarke 1998, Ramesh 1998). It is interesting to note that Gotel and Finkelstein (1994) conclude that a particular concern in improving requirements traceability is the need to facilitate informal communication with those responsible for specifying and detailing requirements. Another evaluation of the traceability challenge reported by Ramesh identifies three factors as influencing the implementation of requirements traceability, namely environmental (tools), organisational (external organisational incentive on individual or internal), and development context (process and practices) (Ramesh 1998).

**Model-based testing** is a large research field within which a wide range of formal models and languages for representing requirements have been suggested (Dias Neto 2007). Defining or modelling the requirements in a formal model or language enables the automatic generation of other development artefacts such as test cases, based on the (modelled) requirements. Similarly to the field of traceability, model-based testing also has issues with practical applicability in industrial development (Nebut 2006, Mohagheghi 2008, Yue 2011). Two exceptions to this is provided by Hasling et al. (2008) and by Nebut et al. (2006) who both report on experiences from applying model-based testing by generating system test cases from UML descriptions of the requirements. The main benefits of model-based testing are in increased test coverage (Nebut 2006, Hasling 2008), enforcing a clear and unambiguous definition of the requirements (Hasling 2008) and increased testing productivity (Grieskamp 2011). However, the formal representation of requirements often results in difficulties both in requiring special competence to produce (Nebut 2006), but also for non-specialist (e.g. business people) in understanding the requirements (Lubars 1993). Transformation of textual requirements into formal models could alleviate some of these issues. However, additional research is required before a practical solution is available for supporting such transformations (Yue 2011). The generation of test cases directly from the requirements implicitly links the two without any need for manually creating (or maintaining) traces. However, depending on the level of the model and the generated test cases the value of the traces might vary. For example, for use cases and system test cases the tracing was reported as being more *natural* than when using state machines (Hasling 2008). Errors in the models are an additional issue to consider when applying model-based testing (Hasling 2008). Scenario-based models where test cases are defined to cover requirements defined as use cases, user stories or user scenarios have been proposed as an alternative to the formal models, e.g. by Regnell and Runeson (1998), Regnell et al. (2000) and Melnik et al. (2006). The scenarios define the requirements at a high level while the details are defined as test cases; acceptance test cases are used to document the detailed requirements. This is an approach often applied in agile development (Cao 2008). Melnik et al. (2006) found that using executable acceptance test cases as detailed requirements is straight-forward to implement and breeds a testing mentality. Similar positive experiences with defining requirements as scenarios and acceptance test cases are reported from industry by Martin et al. (2008)

# 3   Case Study Design

The main goal of this case study was to gain a deeper understanding of the issues in REVV alignment and to identify common practices used in industry to address the challenges within the area. To this end, a flexible exploratory case study design (Robson 2002, Runeson 2012) was chosen with semi-structured interviews as the data collection method. In order to manage the size of the study, we followed a case study process suggested by Runeson et al. (2012, chapter 14) which allowed for a structured approach in managing the large amounts of qualitative data in a consistent manner among the many researchers involved. The process consists of the following five interrelated phases (see Figure 1 for an overview, including in- and outputs of the different phases):
1) *Definition* of goals and research questions
2) *Design and planning* including preparations for interviews
3) *Evidence collection* (performing the interviews)
4) *Data analysis* (transcription, coding, abstraction and grouping, interpretation)
5) *Reporting*

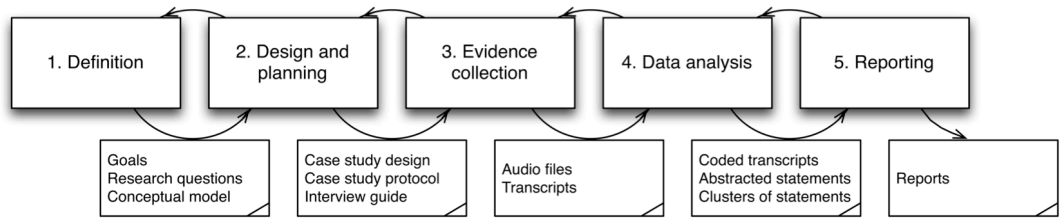

Figure 1. Overview of the research process including in- and output for each phase.

Phases 1-4 are presented in more detail in sections 3.1 to 3.4, while threats to validity are discussed in section 3.5. A more in-depth description with lessons learned from applying the process in this study is presented by Runeson et al (2012, Chapter 14). A description of the six case companies involved in the study can be found in section 3.2.

The ten authors played different roles in the five phases. The senior researchers, Regnell, Gorschek, Runeson and Feldt lead the *goal definition* of the study. They also coached the *design and planning*, which was practically managed by Loconsole, Sabaliauskaite and Engström. *Evidence collection* was distributed over all ten researchers. Loconsole and Sabaliauskaite did the transcription and coding together with Bjarnason, Borg, Engström and Unterkalmsteiner, as well as the preliminary *data analysis* for the evidence from the first company (Sabaliauskaite 2010). Bjarnason, Borg, Engström and Unterkalmsteiner did the major legwork in the intermediate *data analysis*, coached by Regnell, Gorschek and Runeson. Bjarnason and Runeson made the final *data analysis*, *interpretation* and *reporting*, which was then reviewed by the rest of the authors.

## 3.1 Definition of Research Goal and Questions

This initial phase (see Figure 1) provided the direction and scope for the rest of the case study. A set of goals and research questions were defined based on previous experience, results and knowledge of the participating researchers, and a literature study into the area. The study was performed as part of an industrial excellence research centre, where REVV alignment was one theme. Brainstorming sessions were also held with representatives from companies interested in participating in the study. In these meetings the researchers and the company representatives agreed on a main long-term research goal for the area: to improve development efficiency within existing levels of software quality through REVV alignment, where this case study takes a first step into exploring the current state of the art in industry. Furthermore, a number of aspects to be considered were agreed upon, namely agile processes, open source development, software product line engineering, non-functional requirements, and, volume and volatility of requirements. As the study progressed the goals and focal aspects were refined and research questions formulated and documented by two researchers. Four other researchers reviewed their output. Additional research questions were added after performing two pilot interviews (in the next phase, see Section 3.2). In this paper, the following research questions are addressed in the context of software development:

- RQ1: What are the current challenges, or issues, in achieving REVV alignment?
- RQ2: What are the current practices that support achieving REVV alignment?
- RQ3: Which current challenges are addressed by which current practices?

The main concepts of REVV alignment to be used in this study were identified after discussions and a conceptual model of the scope of the study was defined (see Figure 2). This model was based on a traditional V-model showing the artefacts and processes covered by the study, including the relationships between artefacts of varying abstraction level and between processes and artefacts. The discussions undertaken in defining this conceptual model led to a shared understanding within the group of researchers and reduced researcher variation, thus ensuring greater validity of the data collection and results. The model was utilised both as a guide for the researchers in subsequent phases of the study and during the interviews.

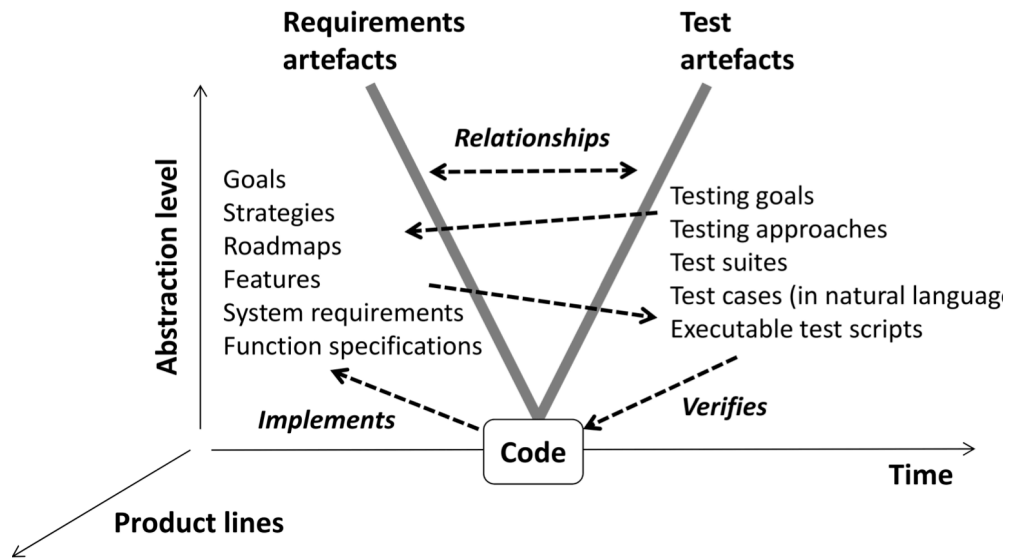

Figure 2. The conceptual model of the area under study, produced in phase 1.

## 3.2 Design and Planning

In this phase, the detailed research procedures for the case study were designed and preparations were made for data collection. These preparations included designing the interview guide and selecting the cases and interviewees.

The interview guide was based on the research questions and aspects, and the conceptual model produced in the *Definition* phase (see Figures 1 and 2). The guide was constructed and refined several times by three researchers and reviewed by another four. User scenarios related to aligning requirements and testing, and examples of alignment metrics were included in the guide as a basis for discussions with the interviewees. The interview questions were mapped to the research questions to ensure that they were all covered. The guide was updated twice; after two pilot interviews, and after six initial interviews. Through these iterations the general content of the guide remained the same, though the structure and order of the interview questions were modified and improved. The resulting interview guide is published by Runeson et al. (2012, appendix C). Furthermore, a consent information letter was prepared to make each interviewee aware of the conditions of the interviews and their rights to refuse to answer and to withdraw at any time. The consent letter is published by Runeson et al. (2012, Appendix E).

The case selection was performed through a brainstorming session held within the group of researchers where companies and interviewee profiles that would match the research goals were discussed. In order to maximise the variation of companies selected from the industrial collaboration network, with respect to size, type of process, application domain and type of product, a combination of maximum variation selection and convenience selection was applied (Runeson 2012, p. 35, 112). The characteristics of the case companies are briefly summarised in Table 1. It is clear from the summary that they represent: a wide range of domains; size from 50 to 1,000 software developers; bespoke and market driven development; waterfall and iterative processes; using open source components or not, etc. At the time of the interviews a major shift in process model, from waterfall to agile, was underway at company F. Hence, for some affected factors in Table 1, information is given as to for which model the data is valid.

Table 1. Overview of the companies covered by this case study. At company F a major process change was taking place at the time of the study and data specific to the previous waterfall-based process are marked with 'previous'.

| Company | A | B | C | D | E | F |
| --- | --- | --- | --- | --- | --- | --- |
| **Type of company** | Software development, embedded products | Consulting | Software development | Systems engineering, embedded products | Software development, embedded products | Software development, embedded products |
| **# employees** | 125-150 | 135 | 500 | 50-100 | 300-350 | 1,000 |
| in software development of targeted organisation | | | | | | |
| # employees in typical project | 10 | Mostly 4-10, but varies greatly | 50-80 | software developers: 10-20 | 6-7 per team, 10-15 teams | Previous process: 800-1,000 person years |
| Distributed | No | Collocated (per project, often on-site at customer) | Yes | Yes | Yes | Yes |
| Domain / System type | Computer networking equipment | Advisory/technical services, application management | Rail traffic management | Automotive | Telecom | Telecom |
| Source of requirements | Market driven | Bespoke | Bespoke | Bespoke | Bespoke and market driven | Bespoke and market driven |
| Main quality focus | Availability, performance, security | Depends on customer focus | Safety | Safety | Availability, Performance, reliability, security | Performance, stability |
| Certification | No software related certification | No | ISO9001, ISO14001, OHSAS18001 | ISO9001, ISO14001 | ISO9001, ISO14001 (aiming towards adhering to TL9000) | ISO9001 |
| Process Model | Iterative | Agile in variants | Waterfall | RUP, Scrum | Scrum, eRUP, a sprints is 3 months | Iterative with gate decisions (agile influenced). Previous: Waterfall |
| Duration of a typical project | 6-18 months | No typical project | 1-5 years to first delivery, then new software release for 1-10 years | 1-5 years to first delivery, then new software releases for 1-10 years | 1 year | Previous process 2 years |
| # requirements in typical project | 100 (20-30 pages HTML) | No typical project | 600-800 at system level | For software: 20-40 use cases | 500-700 user stories | Previous process:14,000 |
| # test cases in a typical project | ~1,000 test cases | No typical project | 250 at system level | | 11,000+ | Previous process 200,000 at platform level, 7,000 at system level |
| Product Lines | Yes | No | Yes | Yes | Yes | Yes |
| Open Source | Yes | Yes. Wide use, including contributions | Yes, partly | No | No | Yes (with new agile process model) |

Our aim was to cover processes and artefacts relevant to REVV alignment for the whole life cycle from requirements definition through development to system testing and maintenance. For this reason, interviewees were selected to represent the relevant range of viewpoints from requirements to testing, both at managerial and at engineering level. Initially, the company contact persons helped us find suitable people to interview. This was complemented by s*nowball sampling* (Robson 2002) by asking the interviewees if they could recommend a person or a role in the company whom we could interview in order to get alignment-related information. These suggestions were then matched against our aim to select interviewees in order to obtain a wide coverage of the processes and artefacts of interest. The selected interviewees represent a variety of roles, working with requirements, testing and development; both engineers and managers were interviewed. The number of interviews per company was selected to allow for going in-depth in one company (company F) through a large number of interviews. Additionally, for this large company the aim was to capture a wide view of the situation and thus mitigate the risk of a skewed sampled. For the other companies, three interviews were held per company. An overview of the interviewees, their roles and level of experience is given in Table 2. Note that for company B, the consultants that were interviewed typically take on a multitude of roles within a project even though they can mainly be characterised as software developers they also take part in requirements analysis and specification, design and testing activities.

Table 2. Overview of interviewees' roles at their companies incl. level of experience in that role;

senior (more than 3 years) or junior (up to 3 years). Xn refers to interviewee n at company X.

Note: most interviewees have additional previous experience.

| Role | A | B | C | D | E | F |
|---|---|---|---|---|---|---|
| Requirements engineer | | | | | | F1 (senior), F6 (senior), F7 (senior) |
| Systems architect | | | | D3 (junior) | E1 (senior) | F4 (senior) |
| Software developer | | B1 (junior), B2 (senior), B3 (senior) | | | | F13 (senior) |
| Test engineer | A2 (senior) | | C1 (senior), C2 (junior) | D2 (senior) | E3 (senior) | F9 (senior), F10 (senior), F11 (junior), F12 (senior), F14 (senior) |
| Project manager | A1 (junior) | | C3 (senior) | D1 (senior) | | F3 (junior), F8 (senior) |
| Product manager | A3 (senior) | | | | E2 (senior) | |
| Process manager | | | | | | F2 (junior), F5 (senior), F15 (junior) |

## 3.3 Evidence Collection

A semi-structured interview strategy (Robson 2002) was used for the interviews, which were performed over a period of one year starting in May 2009. The interview guide (Runeson 2012, appendix C) acted as a checklist to ensure that all selected topics were covered. Interviews lasted for about 90 minutes. Two or three researchers were present at each interview, except for five interviews, which were performed by only one researcher. One of the interviewers led the interview, while the others took notes and asked additional questions for completeness or clarification. After consent was given by the interviewee audio recordings were made of each interview. All interviewees consented.

The audio recordings were transcribed word by word and the transcriptions were validated in two steps to eliminate un-clarities and misunderstandings. These steps were: (i) another researcher, primarily one who was present at the interview, reviewed the transcript, and (ii) the transcript was sent to the interviewee with sections for clarification highlighted and the interviewee had a chance to edit the transcript to correct errors or explain what they meant. These modifications were included into the final version of the transcript, which was used for further data analysis.

The transcripts were divided into chunks of text consisting of a couple of sentences each to enable referencing specific parts of the interviews. Furthermore, an anonymous code was assigned to each interview and the names of the interviewees were removed from the transcripts before data analysis in order to ensure anonymity of the interviewees.

## 3.4 Data Analysis

Once the data was collected through the interviews and transcribed (see Figure 1), a three-stage analysis process was performed consisting of: coding, abstraction and grouping, and interpretation. These multiple steps were required to enable the researchers to efficiently navigate and consistently interpret the huge amounts of qualitative data collected, comprising more than 300 pages of interview transcripts.

**Coding** of the transcripts, i.e. the chunks, was performed to enable locating relevant parts of the large amounts of interview data during analysis. A set of codes, or keywords, based on the research and interview questions was produced, initially at a workshop with the participating researchers. This set was then iteratively updated after exploratory coding and further discussions. In the final version, the codes were grouped into multiple categories at different abstraction levels, and a coding guide was developed. To validate that the researchers performed coding in a uniform way, one interview transcript was selected and coded by all researchers. The differences in coding were then discussed at a workshop and the coding guide was subsequently improved. The final set of codes was applied to all the transcripts. The coding guide and some coding examples are published by Runeson et al. (2012, Appendix D).

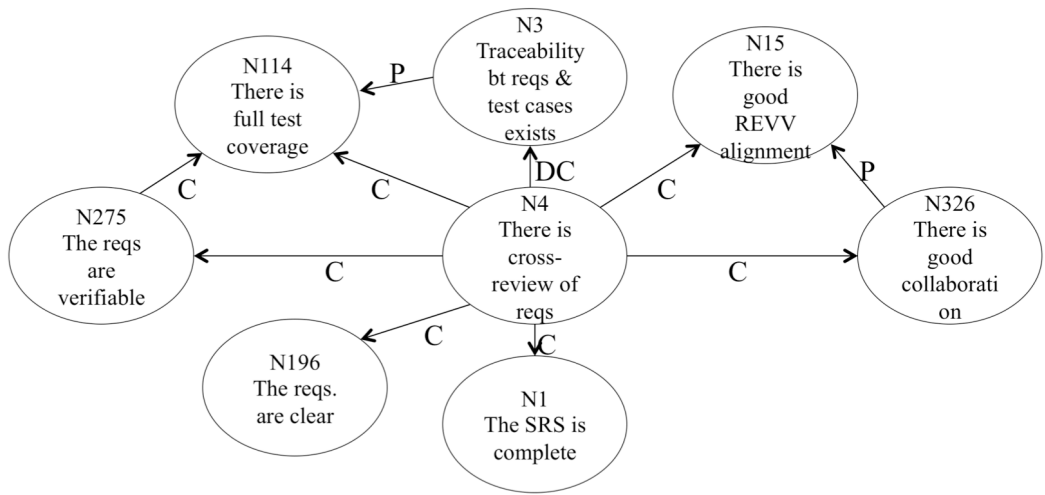

Figure 3.  Part of the abstraction representing the interpretation of the interviewee data. The relationships shown denote C - contribute to, P - prerequisite for, and DC – does not contribute to.

**Abstraction and grouping** of the collected data into statements relevant to the goals and questions for our study was performed in order to obtain a manageable set of data that could more easily be navigated and analysed. The statements can be seen as an index, or common categorisation of sections belonging together, in essence a summary of them as done by Gorschek and Wohlin (2004, 2006), Petterson et al. (2008) and Höst et al. (2010). The statements were each given a unique identifier, title and description. Their relationship to other statements, as derived from the transcripts, was also abstracted. The statements and relationships between them were represented by nodes connected by directional edges. Figure 3 shows an example of the representation designed and used for this study. In particular, the figure shows the abstraction of the interview data around cross-role reviews of requirements, represented by node N4. For example, the statement 'cross-role reviews' was found to contribute to statements related to requirements quality. Each statement is represented by a node. For example, N4 for 'cross-role review', and N1, N196 and N275 for the statements related to requirements quality. The connections between these statements are represented by a 'contributes to' relationship from N4 to each of N1, N196 and N275. These connections are denoted by a directional edge tagged with the type of relationship. For example, the tags 'C' for 'contributes to', 'P' for 'prerequisite for' and 'DC' for 'does not contribute to'. In addition, negation of one or both of the statements can be denoted by applying a post- or prefix 'not' (N) to the connection. The type of relationships used for modelling the connections between statements were discussed, defined and agreed on in a series of work meetings. Traceability to the origin of the statements and the relationships between them was captured and maintained by noting the id of the relevant source chunk, both for nodes and for edges. This is not shown in Figure 3.

The identified statements including relationships to other statements were extracted per transcript by one researcher per interview. To ensure a consistent abstraction among the group of researchers and to enhance completeness and correctness, the abstraction for each interview was reviewed by at least one other researcher and agreed after discussing differences of opinion. The nodes and edges identified by each researcher were merged into one common graph consisting of 341 nodes and 552 edges.

**Interpretation** of the collected evidence involved identifying the parts of the data relevant to a specific research question. The abstracted statements derived in the previous step acted as an index into the interview data and allowed the researchers to identify statements relevant to the research questions of challenges and practices. This interpretation of the interview data was performed by analysing a graphical representation of the abstracted statements including the connections between them. Through the analysis nodes and clusters of nodes related to the research questions were identified. This is similar to explorative coding and, for this paper, the identified codes or clusters represented REVV alignment challenges and practices with one cluster (code) per challenge and per practice. Due to the large amount of data, the analysis and clustering was initially performed on sub-sets of the graphical representation, one for each company. The identified clusters were then iteratively merged into a common set of clusters for the interviews for all companies. For example, for the nodes shown in Figure 3 the statements 'The requirements are clear' (N196) and 'The requirements are verifiable' (N275) were clustered together into the challenge 'Defining clear and verifiable requirements' (challenge Ch3.2, see Section 4.1) based on connections (not shown in the example) to other statements reflecting that this leads to weak alignment.

Even with the abstracted representation of the interview transcripts, the interpretation step is a non-trivial task which requires careful and skilful consideration to identify the nodes relevant to specific research questions. For this reason, the clustering that was performed by Bjarnason was reviewed and agreed with Runeson. Furthermore, the remaining un-clustered nodes were reviewed by Engström, and either mapped to existing clusters, suggested for new clusters or judged to be out of scope for the specific research questions. This mapping was then reviewed and agreed with Bjarnason.

Finally, the agreed clusters were used as an index to locate the relevant parts of the interview transcripts (through traces from the nodes and edges of each cluster to the chunks of text). For each identified challenge and practice, and mapping between them, the located parts of the transcriptions were then analysed and interpreted, and reported in this paper in Sections 4.1, 4.2 and 4.3, respectively for challenges, practices, and the mapping.

## 3.5 Threats to Validity

There are limitations and threats to the validity to all empirical studies, and so also for this case study. As suggested by Runeson et al (2009, 2012), the construct validity, external validity and reliability were analysed in the phases leading up to the *analysis* phase of the case study, see Figure 1. We also report measures taken to improve the validity of the study.

### 3.5.1 Construct Validity

Construct validity refers to how well the chosen research method has captured the concepts under study. There is a risk that academic researchers and industry practitioners may use different terms and have different frames of reference, both between and within these categories of people. In addition, the presence of researchers may threaten the interviewees and lead them to respond according to assumed expectations. The selection of interviewees may also give a limited or unbalanced view of the construct. In order to mitigate these risks, we took the following actions in the design step:

*- Design of the interview guide and reference model.* The interview guide was designed based on the research questions and reviewed for completeness and consistency by other researchers. It was piloted during two interviews and then revised again after another six. The risk that the language and terms used may not be uniformly understood was addressed by producing a conceptual model (see Figure 2), which was shown to the interviewees to explain the terminology. However, due to the semi-structured nature of the guide and the different interviewers involved the absence of interviewee data for a certain concept, challenge or practice cannot be interpreted as the absence of this item either in the interviewees experience or in the company. For similar reasons, the results do not include any ranking or prioritisation as to which challenges and practices are the most frequent or most effective.

- *Prolonged involvement*. The companies were selected so that at least one of the researchers had a long-term relation with them. This relationship helped provide the trust needed for openness and honesty in the interviews. To mitigate the bias of knowing the company too well, all but five interviews (companies D and E) were conducted by more than one interviewer.

- *Selection of interviewees*. To obtain a good representation of different aspects, a range of roles were selected to cover requirement, development and testing, and also engineers as well as managers, as reported in Table 2. The aim was to cover the relevant aspects described in the conceptual model, produced during the *Definition* phase (see Section 3.1, Figures 1 and 2). There is a risk that the results might be biased due to a majority of the interviewees being from Company F. However, the results indicate that this risk was minor, since a majority of the identified items (see Section 4) could be connected to multiple companies.

- *Reactive bias*: The presence of a researcher might limit or influence the outcome either by hiding facts or responding after assumed expectations. To reduce this threat the interviewees were guaranteed anonymity both within the company and externally. In addition, they were not given any rewards for their participation and had the right to withdraw at any time without requiring an explanation, though no interviewees did withdraw. This approach indicated that we were interested in obtaining a true image of their reality and encouraged the interviewees to share this.

*3.5.2 Internal Validity*

Even though the conclusions in this paper are not primarily about causal relations, the identification of challenges and practices somewhat resembles identifying factors in casual relations. In order to mitigate the risk of identifying incorrect factors, we used data source triangulation by interviewing multiple roles at a company. Furthermore, extensive observer triangulation was applied in the analysis by always including more than one researcher in each step. This strategy also partly addressed the risk of incorrect generalisations when abstracting challenges and practices for the whole set of companies. However, the presented results represent one possible categorisation of the identified challenges and practices. This is partly illustrated by the fact that not all identified practices can be connected to a challenge.

The interviews at one of the case companies were complicated by a major process change that was underway at the time of the study. This change posed a risk of confusing the context for which a statement had been experienced; the previous (old) way of working or the newly introduced agile practices. To mitigate this risk, we ensured that we correctly understood which process the response concerned, i.e. the previous or the current process.

Furthermore, due to the nature of semi-structured interviews in combination with several different interviewers it is likely that different follow-on questions were explored by the various researchers. This risk was partly mitigated by jointly defining the conceptual model and agreeing on a common interview guide that was used for all interviews. However, the fact remains that there are differences in the detailed avenues of questioning which has resulted in only being able to draw conclusions concerning what was actually said at the interviews. So, for example, if the completeness of the requirements specification (Ch3.2) was not explicitly discussed at an interview no conclusions can be drawn concerning if this is a challenge or not for that specific case.

*3.5.3 External Validity*

For a qualitative study like this, external validity can never be assured by sampling logic and statistical generalisation, but by analytical generalisation which enables drawing conclusions and, under certain conditions, relating them also to other cases (Robson 2002, Runeson 2012). This implies that the context of the study must be compared to the context of interest for the findings to be generalised to. To enable this process, we report the characteristics of the companies in as much detail as possible considering confidentiality (see Table 1). The fact that six different companies of varying size and domain are covered by the study, and some results are connected to the variations between them indicates that the results are more general than if only one company had been studied. But, of course, the world consists of more than six kinds of companies, and any application of the results of this study need to be mindfully tailored to other contexts.

*3.5.4 Reliability*

The reliability of the study relates to whether the same outcome could be expected with another set of researchers. For qualitative data and analysis, which are less procedural than quantitative methods, exact replication is not probable. The analysis lies in interpretation and coding of words, and the set of codes would probably be partly different with a different set of researchers.

To increase the reliability of this study and to reduce the influence by single researchers, several researchers have taken part in the study in different roles. All findings and each step of analysis have been reviewed by and agreed with at least one other researcher. In addition, a systematic and documented research process has been applied (see Figure 1) and a trace of evidence has been retained for each analysis steps. The traceability back to each source of evidence is documented and kept even in this report to enable external assessment of the chain of evidence, if confidentially agreements would allow.

Finally, the presentation of the findings could vary depending on categorisation of the items partly due to variation in views and experience of individual researchers. For example, a challenge in achieving alignment such as Ch2 *Collaborating successfully* (see Section 4.1.2) could be identified also as a practice at the general level, e.g. *to collaborate successfully* could be defined as an alignment practice. However, we have chosen to report specific practices that may improve collaboration and thereby REVV alignment. For example, P1.1 *Customer communication at all requirements levels and phases* can support improved coordination of requirements between the customer and the development team. To reduce the risk of bias in this aspect, the results and the categorisation of them was first proposed by one researcher and then reviewed by four other researchers leading to modifications and adjustments.

# 4 Results

Practitioners from all six companies in the study found alignment of RE with VV to be an important, but challenging, factor in developing products. REVV alignment was seen to affect the whole project life cycle, from the contact with the customer and throughout software development. The interviewees stated clearly that good alignment is essential to enable smooth and efficient software development. It was also seen as an important contributing factor in producing software that meets the needs and expectations of the customers. A software developer stated that alignment is 'very important in creating the right system' (B1:27[1]). One interviewee described the customer's view of a product developed with misaligned requirements as: 'There wasn't a bug, but the behaviour of the functionality was interpreted or implemented in such a way that it was hard to do what the customer [originally] intended.' (A3:43) Another interviewee mentioned that alignment between requirements and verification builds customer trust in the end product since good alignment allows the company to 'look into the customer's eyes and explain what have we tested… on which requirements' (D2:10).

In general, the interviewees expressed that weak and unaligned communication of the requirements often cause inconsistencies that affect the verification effort. A common view was that these inconsistencies, caused by requirements that are misunderstood, incorrect or changed, or even un-communicated, leads to additional work in updating and re-executing test cases. Improved alignment, on the other hand, was seen to make 'communication between different levels in the V-model a lot easier' (E3:93). One of the interviewed testers stated: 'Alignment is necessary. Without it we [testers] couldn't do our job at all.' (C1:77)

Below, we present the results concerning the challenges of alignment (Ch1-Ch10) and the practices (P1-P10) used, or suggested, by the case companies to address REVV challenges. Table 3 provides an overview of the challenges found for each company, while Table 4 contains an overview of the practices. Table 6 shows which challenges each practices is seen to address.

## 4.1 Alignment Challenges

The alignment challenges identified through this study are summarised in Table 3. Some items have been categorised together as one challenge, resulting in 10 main challenges where some consist of several related challenges. For example, Ch3 *Requirements specification quality* consists of three challenges (Ch3.1-Ch3.3) concerning different aspects of requirements quality. Each challenge including sub items is described in the subsections that follow.

Table 3. Alignment challenges mentioned for each company. Note: a blank cell means that the challenge was not mentioned during the interviews, not that it is not experienced.

| | Id | Challenge | Company | | | | | |
|---|---|---|---|---|---|---|---|---|
| | | | A | B | C | D | E | F |

---

[1] Reference to source is given by interviewee code, see Table 2.

|  | Ch1 | Aligning goals and perspectives within an organisation | X | X | X |  | X | X |
| --- | --- | --- | --- | --- | --- | --- | --- | --- |
|  | Ch2 | Cooperating successfully | X |  | X | X | X | X |
| Req spec quality | Ch3.1 | Defining clear and verifiable requirements |  |  | X | X | X | X |
|  | Ch3.2 | Defining complete requirements |  | X |  | X | X | X |
|  | Ch3.3 | Keeping requirements documents updated |  |  |  |  |  | X |
| VV quality | Ch4.1 | Full test coverage | X | X | X | X |  | X |
|  | Ch4.2 | Defining a good verification process |  |  |  |  |  | X |
|  | Ch4.3 | Verifying quality requirements |  | X |  | X |  | X |
|  | Ch5 | Maintaining alignment when requirements change | X |  | X |  |  | X |
| Req's abstract levels | Ch6.1 | Defining requirements at abstraction level well matched to test cases |  |  |  | X |  | X |
|  | Ch6.2 | Coordinating requirements at different abstraction levels | X |  |  |  |  | X |
| Traceability | Ch7.1 | Tracing between requirements and test cases | X | X | X | X |  | X |
|  | Ch7.2 | Tracing between requirements abstraction levels |  | X | X | X |  |  |
|  | Ch8 | Time and resource availability |  |  | X |  | X | X |
|  | Ch9 | Managing a large document space |  |  | X | X |  | X |
|  | Ch10 | Outsourcing of components or testing |  |  |  | X |  | X |

*4.1.1 Challenge 1: Aligning Goals and Perspectives within an Organisation (Ch1)*
The alignment of *goals* throughout the organisation was mentioned by many interviewees as vital in enabling cooperation among different organisational units (see challenge 2 in Section 4.1.2). However, goals were often felt to be missing or unclearly defined, which could result in 'making it difficult to test [the goals]' (B3:17). In several companies problems with differing and unaligned goals were seen to affect the synchronisation between requirements and testing, and cause organisational units to counteract each other in joint development projects. For example, a product manager mentioned that at times, requirement changes needed from a business perspective conflicted with the goals of the development units; 'They [business roles] have their own directives and … schedule target goals' and 'they can look back and see which product was late and which product was good' (A3:74). In other words, misaligned goals may have an impact on both time schedules and product quality.

Many interviewees described how awareness and understanding of different *perspectives* on the problem domain is connected to better communication and cooperation, both towards the customers and external suppliers, and internally between competence areas and units. When there is a lack of aligned perspectives, the customer and the supplier often do not have the same understanding of the requirements. This may result in 'errors in misunderstanding the requirements' (B3:70). Lack of insight into and awareness of different perspectives was also seen to result in decisions (often made by other units) being questioned and requirements changed at a late stage in the development cycle with a subsequent increase in cost and risk. For example, a systems architect described that in a project where there is a 'higher expectations on the product than we [systems architect] scoped into it' (E1:20) a lot of issues and change requests surface in the late project phases. A software developer stated concerning the communication between requirements engineers and developers that 'if both have a common perspective [of technical possibilities], then it would be easier to understand what [requirements] can be set and what cannot be set' (F13:29). Or in other words, with an increased common understanding technically infeasible requirements can be avoided already at an early stage.

Weak alignment of goals and perspectives implies a weak coordination at higher organisational levels and that strategies and processes are not synchronised. As stated by a process manager, the involvement of many separate parts of an organisation then leads to 'misunderstandings and misconceptions and the use of different vocabulary' (F2:57). In addition, a test engineer at Company A mentioned that for the higher abstraction levels there were no attempts to synchronise, for example, the testing strategy with the goals of development projects to agree on important areas to focus on (A2:105). Low maturity of the organisation was thought to contribute to this and result in the final product having a low degree of correspondence to the high-level project goals. A test engineer said: 'In the long run, we would like to get to the point where this [product requirements level] is aligned with this [testing activities].' (A2:119)

*4.1.2 Challenge 2: Cooperating Successfully (Ch2)*

All of the companies included in our study described close cooperation between roles and organisational units as vital for good alignment and coordination of both people and artefacts. Weak cooperation is experienced to negatively affect the alignment, in particular at the product level. A product manager stated that 'an "us and them" validation of product level requirements is a big problem' (A3:058-059). Ensuring clear agreement and communication concerning which requirements to support is an important collaboration aspect for the validation. At Company F (F12:063) lack of cooperation in the early phases in validating requirements has been experienced to result in late discovery of failures in meeting important product requirements. The development project then say at a late stage: 'We did not approve these requirements, we can't solve it' (F12:63) with the consequence that the requirements analysis has to be re-done. For Company B (consulting in different organisations) cooperation and communication was even described as being prioritised above formal documentation and processes, expressed as: 'We have succeeded with mapping requirements to tests since our process is more of a discussion' (B3:49). Several interviewees described that alignment at product and system level, in particular, is affected by how well people cooperate (C2:17, E1:44, 48, E2:48, F4:66, F15:46). When testers have a good cooperation and frequently communicate with both requirements-related and development-related roles, this leads to increased alignment (E3:093).

Organisational boundaries were mentioned as further complicating and hindering cooperation between people for two of the companies, namely companies E and F. In these cases, separate organisational units exist for requirements (E2:29, E3:94, F2:119), usability (F10:108) and testing (F3:184). As one interviewee said: 'it is totally different organisations, which results in ... misunderstandings and misconceptions...we use different words' (F02:57). Low awareness of the responsibilities and tasks of different organisational units was also claimed to negatively affect alignment (F2:264). This may result in increased lead times (E1:044, F15:033), need for additional rework (E1:150, E1:152), and conflicts in resource allocation between projects (F10:109, E1:34).

*4.1.3 Challenge 3: Good Requirements Specification Quality (Ch3)*

'If we don't have good requirements the tests will not be that good.' (D3:14) When the requirement specification is lacking the testers need to guess and make up the missing information since 'the requirements are not enough for writing the software and testing the software' (D3:19). This both increases the effort required for testing and the risk of misinterpretation and missing vital customer requirements. One process manager expressed that the testability of requirements can be improved by involving testers and that 'one main benefit [of alignment] is improving the requirements specifications' (F2:62). A test leader at the same company identified that a well aligned requirements specification (through clear agreement between roles and tracing between artefacts) had positive effects such as 'it was very easy to report when we found defects, and there were not a lot of discussions between testers and developers, because everyone knew what was expected' (F9:11).

There are several aspects to good requirements that were found to relate to alignment. In the study, practitioners mentioned good requirements as being verifiable, clear, complete, at the right level of abstraction, and up-to-date. Each aspect is addressed below.

- **Defining clear and verifiable requirements (Ch3.1)** was mentioned as a major challenge in enabling good alignment of requirements and testing, both at product and at detailed level. This was mentioned for four of the six companies covered by our study, see Table 3. Unclear and non-verifiable requirements were seen as resulting in increased lead times and additional work in later phases in clarifying and redoing work based on unclear requirements (F2:64, D1:80). One test manager said that 'in the beginning the requirements are very fuzzy. So it takes time. And sometimes they are not happy with our implementation, and we have to do it again and iterate until it's ready.' (F11:27, similar in E3:44.) Failure to address this challenge ultimately results in failure to meet the customer expectations with the final product. A project manager from company D expressed this by saying that non-verifiable requirements is the reason 'why so many companies, developers and teams have problems with developing customer-correct software' (D1:36).

- **Defining complete requirements (Ch3.2)** was claimed to be required for successful alignment by interviewees from four companies, namely companies B, D, E and F. As expressed by a systems architect from Company D, 'the problem for us right now is not [alignment] between requirements and testing, but that the requirements are not correct and complete all the time' (D3:118). Complete requirements support achieving full test coverage to ensure that the full functionality and quality aspects are verified. (F14:31) When testers are required to work with incomplete requirements, additional information is acquired from other sources, which requires additional time and effort to locate (D3:19).

- **Keeping requirements documentation updated (Ch3.3)** Several interviewees from company F described how a high frequency of change leads to the requirements documentation not being kept updated, and consequently the documentation cannot be relied on (F14:44, F5:88). When a test for a requirement then fails, the first reaction is not: 'this is an error', but rather 'is this really a relevant requirement or should we change it' (F5:81). Mentioned consequences of this include additional work to locate and agree to the correct version of requirements and rework (F3:168) when incorrect requirements have been used for testing. Two sources of requirements changes were mentioned, namely requested changes that are formally approved (F14:50), but also changes that occur as the development process progresses (during design, development etc.) that are not raised as formal change requests (F5:82, F5:91, F11:38). When the requirements documentation is not reliable, the projects depend on individuals for correct requirements information. As expressed by one requirements engineer: 'when you lose the people who have been involved, it is tough. And, things then take more time.' (F1:137)

*4.1.4 Challenge 4: Validation and Verification Quality*

Several issues with validation and verification were mentioned as alignment challenges that affect the efficiency and effectiveness of the testing effort. One process manager with long experience as a tester said: 'We can run 100,000 test cases but only 9% of them are relevant.' (F15:152) Testing issues mentioned as affecting alignment were: obtaining full test coverage, having a formally defined verification process and the verification of quality requirements.

- **Full test coverage (Ch4.1)** Several interviewees described full test coverage of the requirements as an important aspect of ensuring that the final product fulfils the requirements and the expectations of the customers. As one software developer said: 'having full test coverage with unit tests gives a better security... check that I have interpreted things correctly with acceptance tests' (B1:117). However, as a project manager from Company C said: 'it is very hard to test everything, to think about all the complexities' (C3:15). *Unclear* (Ch3.2, C1:4) and *non-verifiable requirements* (Ch3.1, A1:55, D1:78, E1:65) were mentioned as contributing to difficulties in achieving full test coverage of requirements for companies A, B, D and E. For certain requirements that are expressed in a verifiable way a project manager mentioned that they cannot be tested due to limitations in the process, competence and test tools and environments (A1:56). To ensure full test coverage of requirements the testers need knowledge of the full set of requirements, which is impeded in the case of *incomplete requirements specifications* (Ch3.3) where features and functionality are not described (D3:16). This can also be the case for *requirements defined at a higher abstraction level* (F2:211, F14:056). *Lack of traceability* between requirements and test cases was stated to making it harder to know when full *test coverage* has been obtained (A1:42). For company C, traceability was stated as time consuming but necessary to ensure and demonstrate full test coverage, which is mandatory when producing safety-critical software (C1:6, C1:31). Furthermore, obtaining sufficient coverage of the requirements requires analysis of both the requirement and the connected test cases (C1:52, D3:84, F14:212). As one requirements engineer said, 'a test case may cover part of a requirement, but not test the whole requirement' (F7:52). Late requirements changes was mentioned as a factor contributing to the challenge of full test coverage (C1:54, F7:51) due to the need to update the affected test cases, which is hampered by failure to keep the requirements specification updated after changes (Ch3.5, A2:72, F15:152).
- **Having a verification process (Ch4.2)** was mentioned as directly connected to good alignment between requirements and test. At company F, the on-going shift towards a more agile development process had resulted in the verification unit operating without a formal process (F15:21). Instead each department and project 'tries to work their own way... that turns out to not be so efficient' (F15:23), especially so in this large organisation where many different units and roles are involved from the initial requirements definition to the final verification and launch. Furthermore, one interviewee who was responsible for defining the new verification process (F15) said that 'the hardest thing [with defining a process] is that there are so many managers ... [that don't] know what happens one level down'. In other words, a verification process that supports requirements-test alignment needs to be agreed with the whole organisation and at all levels.

- **Verifying quality requirements (Ch4.3)** was mentioned as a challenge for companies B, D and F. Company B has verification of quality in focus with continuous monitoring of quality levels in combination with frequent releases; 'it is easy to prioritise performance optimisation in the next production release' (B1:52). However, they do not work proactively with quality requirements. Even though they have (undocumented) high-level quality goals the testers are not asked to use them (B1:57, B2:98); 'when it's not a broken-down [quality] requirement, then it's not a focus for us [test and development]' (B3:47). Company F does define formal quality requirements, but these are often not fully agreed with development (F12:61). Instead, when the specified quality levels are not reached, the requirements, rather than the implementation, are changed to match the current behaviour, thus resigning from improving quality levels in the software. As one test engineer said: 'We currently have 22 requirements, and they always fail, but we can't fix it' (F12:61). Furthermore, defining verifiable quality requirements and test cases was mentioned as challenging, especially for usability requirements (D3:84, F10:119). Verification is then faced with the challenge of subjectively judging if a requirement is passed or failed (F2:46, F10:119). At company F, the new agile practices of detailing requirements at the development level together with testers was believed to, at least partly, address this challenge (F12:65). Furthermore, additional complication is that some quality requirements can only be verified through analysis and not through functional tests (D3:84).

*4.1.5   Challenge 5: Maintaining Alignment when Requirements Change (Ch5)*

Most of the companies of our study face the challenge of maintaining alignment between requirements and tests as requirements change. This entails ensuring that both artefacts and tracing between them are updated in a consistent manner. Company B noted that the impact of changes is specifically challenging for test since test code is more sensitive to changes than requirements specifications. 'That's clearly a challenge, because [the test code is] rigid, as you are exemplifying things in more detail. If you change something fundamental, there are many tests and requirements that need to be modified' (B3:72).

Loss of traces from test cases to requirements over time was also mentioned to cause problems. When test cases for which traces have been outdated or lost are questioned, then 'we have no validity to refer to ... so we have to investigate' (A2:53). In company A, the connection between requirements and test cases are set up for each project (A2:71): 'This is a document that dies with the project'; a practice found very inefficient. Other companies had varying ambitions of a continuous maintenance of alignment and traces between the artefacts. A key for maintaining alignment when requirements change is that the requirements are actively used. When this is not the case there is a need for obtaining requirements information from other sources. This imposes a risk that 'a requirement may have changed, but the software developers are not aware of it' (D3:97).

Interviewees implicitly connected the traceability challenge to tools, although admitting that 'a tool does not solve everything... Somebody has to be responsible for maintaining it and to check all the links ... if the requirements change' (C3:053). With or without feasible tools, tracing also requires personal assistance. One test engineer said, 'I go and talk to him and he points me towards somebody' (A2:195).

Furthermore, the *frequency of changes* greatly affects the extent of this challenge and is an issue when trying to establish a base-lined version of the requirements. Company C has good tool support and traceability links, but require defined versions to relate changes to. In addition, they have a product line, which implies that the changes must also be coordinated between the platform (product line) and the applications (products) (C3:019, C3:039).

*4.1.6   Challenge 6: Requirements Abstraction Levels (Ch6)*

REVV alignment was described to be affected by the abstraction levels of the requirements for companies A, D and F. This includes the relationship to the abstraction levels of the test artefacts and ensuring consistency between requirements at different abstraction levels.

- **Defining requirements at abstraction levels well-matched to test cases (Ch6.1)** supports defining test cases in line with the requirements and with a good coverage of them. This was mentioned for companies D and F. A specific case of this at company D is when the testers 'don't want to test the complete electronics and software system, but only one piece of the software' (D3:56). Since the requirements are specified at a higher abstraction level than the individual components, the requirements for this level then need to be identified elsewhere. Sources for information mentioned by the interviewees include the design specification, asking people or making up the missing requirements (D3:14).

This is also an issue when retesting only parts of a system which are described by a high-level requirement to which many other test cases are also traced (D3:56). Furthermore, synchronising the abstraction levels between requirements and test artefacts was mentioned to enhance coverage (F14:31).

- **Coordinating requirements at different abstraction levels (Ch6.2)** when breaking down the high-level requirements (such as goals and product concepts) into detailed requirements at system or component level was mentioned as a challenge by several companies. A product manager described that failure to coordinate the detailed requirements with the overall concepts could result in that 'the intention that we wanted to fulfil is not solved even though all the requirements are delivered' (A3:39). On the other hand, interviewees also described that the high-level requirements were often vague at the beginning when 'it is very difficult to see the whole picture' (F12:144) and that some features are 'too complex to get everything right from the beginning' (A3:177).

### 4.1.7 Challenge 7: Tracing between Artefacts (Ch7)

This challenge covers the difficulties involved in tracing requirements to test cases, and vice versa, as well as, tracing between requirements at different abstraction levels. Specific tracing practices identified through our study are described in Sections 4.2.6 and 4.2.7.

- **Tracing between requirements and test cases (Ch7.1).** The most basic kind of traceability, referred to as 'conceptual mapping' in Company A (A2:102), is having a line of thought (not necessarily documented) from the requirements through to the defining and assessing of the test cases. This cannot be taken for granted. Lack of this basic level of tracing is largely due to weak awareness of the role requirements in the development process. As a requirements process engineer in Company F says, 'One challenge is to get people to understand why requirements are important; to actually work with requirements, and not just go off and develop and do test cases which people usually like doing' (F5:13).

    Tracing by using matrices to map between requirements and test cases is a major cost issue. A test architect at company F states, that 'we don't want to do that one to one mapping all the way because that takes a lot of time and resources' (F10:258). Companies with customer or market demands on traceability, e.g. for safety critical systems (companies C and D), have full traceability in place though 'there is a lot of administration in that, but it has to be done' (C1:06). However, for the other case companies in our study (B3:18, D3:45; E2:83; F01:57), it is a challenge to implement and maintain this support even though tracing is generally seen as supporting alignment. Company A says 'in reality we don't have the connections' (A2:102) and for Company F 'in most cases there is no connection between test cases and requirements' (F1:157). Furthermore, introducing traceability may be costly due to large legacies (F1:57) and maintaining traceability is costly. However, there is also a cost for lack of traceability. This was stated by a test engineer in Company F who commented on degrading traceability practices with 'it was harder to find a requirement. And if you can't find a requirement, sometimes we end up in a phase where we start guessing' (F12:112).

    Company E has previously had a tradition of 'high requirements on the traceability on the products backwards to the requirements' (E2:83). However, this company foresees problems with the traceability when transitioning towards agile working practices, and using user stories instead of traditional requirements. A similar situation is described for Company F, where they attempt to solve this issue by making the test cases and requirements one; 'in the new [agile] way of working we will have the test cases as the requirements' (F12:109).

    Finally, traceability for quality (a.k.a. non-functional) requirements creates certain challenges, 'for instance, for reliability requirement you might ... verify it using analysis' (D3:84) rather than testing. Consequently, there is no single test case to trace such a quality requirement to, instead verification outcome is provided through an analysis report. In addition, tracing between requirements and test cases is more difficult 'the higher you get' (B3:20). If the requirements are at a high abstraction level, it is a challenge to define and trace test cases to cover the requirements.

- **Tracing between requirements abstraction levels (Ch7.2)** Another dimension of traceability is vertical tracing between requirements at different abstraction levels. Company C operates with a detailed requirements specification, which for some parts consists of sub-system requirements specifications (C1:31). In this case, there are no special means for vertical traceability, but pointers in the text. It is similar in Company D, where a system architect states that 'sometimes it's not done on each individual requirement but only on maybe a heading level or something like that' (D3:45). Company F use a high-end requirements management tool, which according to the requirements engineer 'can trace the requirement from top level to the lowest implementation level' (F7:50).

Company E has requirements specifications for different target groups, and hence different content; one market oriented, one product oriented, and one with technical details (E1:104). The interviewee describes tracing as a 'synch activity' without specifying in more detail. Similarly, Company F has 'roadmaps' for the long term development strategy, and there is a loosely coupled 'connection between the roadmaps and the requirements' to balance the project scope against strategy and capacity (F11:50).

*4.1.8 Challenge 8: Time and Resource Availability (Ch8)*

In addition to the time consuming task of defining and maintaining traces (Ch7) further issues related to time and resources were brought forward in companies C, E and F. Without sufficient resources for validation and verification the amount of testing that can be performed is not sufficient for the demands on functionality and quality levels expected of the products. The challenge of planning for *enough test resources* is related to the alignment between the proposed requirements and the time and resources required to sufficiently test them. A requirements engineer states that 'I would not imagine that those who are writing the requirements in anyway are considering the test implications or the test effort required to verify them' (F6:181). A test manager confirms this view (F14:188). It is not only a matter of the amount of resources, but also in which time frame they are available (E1:18). Furthermore, availability of all the necessary *competences and skills* within a team was also mentioned as an important aspect of ensuring alignment. A software developer phrased it: 'If we have this kind of people, we can set up a team that can do that, and then the requirements would be produced properly and hopefully 100% achievable' (F13:149). In addition, experienced individuals were stated to contribute to strengthening the alignment between requirements and testing, by being 'very good at knowing what needs to be tested and what has a lower priority' (C2:91), thereby increasing the test efficiency. In contrast, inexperienced testing teams were mentioned for Company C as contributing to weaker alignment towards the overall set of requirements including goals and strategies since they 'verify only the customer requirements, but sometimes we have hazards in the system which require the product to be tested in a better way' (C2:32-33).

*4.1.9 Challenge 9: Managing a Large Document Space (Ch9)*

The main challenge regarding the information management problems lies in the sheer numbers. A test engineer at Company F estimates that they have accumulated 50,000 requirements in their database. In addition, they have 'probably hundreds of thousands of test cases' (F2:34, F12:74). Another test engineer at the same company points out that this leads to information being *redundant* (F11:125), which consequently may lead to inconsistencies. A test engineer at Company D identifies the *constant change* of information as a challenge; they have difficulties to work against the same baseline (D2:16).

Another test engineer at Company F sees information management as a tool issue. He states that 'the requirements tool we have at the moment is not easy to work with....Even, if they find the requirements they are not sure they found the right version' (F9:81). In contrast, a test engineer at company C is satisfied with the ability to find information in the same tool (C2). A main difference is that at Company F, 20 times as many requirements are handled than at Company C.

The investment into introducing explicit links between a huge legacy of requirements and test cases is also put forward as a major challenge for companies A and F. In addition, connecting and integrating different tools was also mentioned as challenging due to separate responsibilities and competences for the two areas of requirements and testing (F5:95, 120).

*4.1.10 Challenge 10: Outsourcing or Offshoring of Components or Testing (Ch10)*

Outsourcing and offshoring of component development and testing create challenges both in agreeing to which detailed requirements to implement and test, and in tracing between artefacts produced by different parties. Company D stresses that the *timing* of the outsourcing plays a role in the difficulties in tracing component requirement specifications to the internal requirements at the higher level; 'I think that's because these outsourcing deals often have to take place really early in the development.' (D3:92). Company F also mentions the timing aspect for acquisition of hardware components; 'it is a rather formal structured process, with well-defined deliverables that are slotted in time' (F6:21).

When testing is outsourced, the *specification* of what to test is central and related to the type of testing. The set-up may vary depending on competence or cultural differences etc. For example, company F experienced that cultural aspects influence the required level of detail in the specification; 'we [in Europe] might have three steps in our test cases, while the same test case with the same result, but produced in China, has eight steps at a more detailed level' (F15:179). A specification of what to test may be at a high level and based on a requirements specification from which the in-sourced party derives tests and executes. An alternative approach is when a detailed test specification is requested to be executed by the in-sourced party (F6:251-255).

## 4.2 Practices for Improved Alignment

This study has identified 27 different alignment practices, grouped into 10 categories. Most of the practices are applied at the case companies, though some are suggestions made by the interviewees. These categories and the practices are presented below and discussed and summarised in Section 5. In Section 4.3 they are mapped to the challenges that they are seen to address.

Table 4. Alignment practices and categories, and case companies for which they were mentioned. Experienced practices are marked with X, while suggested practices are denoted with S. Note: a blank cell means that the practice was not mentioned during the interviews. It does not mean that it is not applied at the company.

| Cat. | Id | Description | A | B | C | D | E | F |
|---|---|---|---|---|---|---|---|---|
| Requirements | P1.1 | Customer communication at all requirements levels and phases | | X | X | X | X | X |
| | P1.2 | Development involved in detailing requirements | X | X | | | | X |
| | P1.3 | Cross-role requirements reviews | X | | X | X | X | X |
| | P1.4 | Requirements review responsibilities defined | | | | | X | X |
| | P1.5 | Subsystem expert involved in requirements definition | | | | X | | X |
| | P1.6 | Documentation of requirement decision rationales | | | | | S | S |
| Validation | P2.1 | Test cases reviewed against requirements | | | | | | X |
| | P2.2 | Acceptance test cases defined by customer | | X | | | | |
| | P2.3 | Product manager reviews prototypes | X | | | | X | |
| | P2.4 | Management base launch decision on test report | | | | | | X |
| | P2.5 | User / Customer testing | | X | | X | X | X |
| Verification | P3.1 | Early verification start | | | | | X | X |
| | P3.2 | Independent testing | | | X | X | X | |
| | P3.3 | Testers re-use customer feedback from previous projects | | | | X | X | X |
| | P3.4 | Training off-shore testers | | | | X | | |
| Change | P4.1 | Process for requirements changes involving VV | X | | X | X | X | X |
| | P4.2 | Product-line requirements practices | X | | X | | | |
| | P5 | Process enforcement | | | | X | | S |
| Tracing | P6.1 | Document-level traces | X | | | | | |
| | P6.2 | Requirements-test case traces | | | | | | X |
| | P6.3 | Test cases as requirements | X | | | | | X |
| | P6.4 | Same abstraction levels for requirements and test spec | | | | X | X | |
| | P7 | Traceability responsibility role | | | | X | X | X |
| Tools | P8.1 | Tool support for requirements and testing | X | | X | X | X | X |
| | P8.2 | Tool support for requirements-test case tracing | X | | X | X | X | X |
| | P9 | Alignment metrics, e.g. test coverage | | | X | X | X | X |
| | P10 | Job rotation | | | | S | | S |

*4.2.1 Requirements Engineering Practices*

Requirements engineering practices are at the core of aligning requirements and testing. This category of practices includes customer communication and involving development-near roles in the requirements process. The interviewees described close cooperation and team work as a way to improve RE practices (F12:146) and thereby the coordination with developers and testers and avoid a situation where product managers say '"redo it" when they see the final product' (F12:143).

- **Customer communication at all levels and in all phases of development (P1.1)** was mentioned as an alignment practice for all but one of the case companies. The communication may take the form of customer-supplier co-location; interaction with the customer based on executable software used for demonstrations or customer validation; or agreed acceptance criteria between customer and supplier. For the smaller companies, and especially those with bespoke requirements (companies B and C), this interaction is directly with a physical customer. In larger companies (companies E and F), and especially within market driven development, a *customer proxy* may be used instead of the real customer, since there is no assigned customer at the time of development or there is a large organisational distance to the customer. Company F assigns a person in each development team 'responsible for the feature scope. That person is to be available all through development and to the validation of that feature' (F2:109). Furthermore, early discussions about product roadmaps from a four to five year perspective are held with customers and key suppliers (F6:29) as an initial phase of the requirements process.
- **Involving developers and testers in detailing requirements (P1.2)** is another practice, especially mentioned by companies A and F. A product manager has established this as a deliberate strategy by conveying the vision of the product to the engineers rather than detailed requirements: 'I'm trying to be more conceptual in my ordering, trying to say what's important and the main behaviour.' (A3:51) The responsibility for detailing the specification then shifts to the development organisation. However, if there is a weak awareness of the customer or market perspectives, this may be a risky practice as 'some people will not [understand this] either because they [don't] have the background or understanding of how customers or end-users or system integrators think' (A3:47). Testers may be involved to ensure the testability of the requirements, or even specify requirements in the form of test cases. Company F was in the process of transferring from a requirements-driven organisation to a design-driven one. Splitting up the (previous) centralised requirements department resulted in 'requirements are vaguer now. So it's more up to the developers and the testers to make their own requirements.' (F12:17) Close cooperation around requirements when working in an agile fashion was mentioned as vital by a product manager from Company E: 'Working agile requires that they [requirements, development, and test] are really involved [in requirements work] and not only review.' (E2:083)
- **Cross-role requirements reviews (P1.3)** across requirements engineers and testers is another practice applied to ensure that requirements are understood and testable (A2:65, C3:69, F2:38, F7:7). The practical procedures for the reviews, however, tend to vary. Company A has an early review of requirements by testers while companies C and D review the requirements while creating the test cases. Different interviewees from companies E and F mentioned one or the other of these approaches; the process seems to prescribe cross-role reviews but process compliance varies. A test engineer said '[the requirements are] usually reviewed by the testers. It is what the process says.' (F11:107) Most interviewees mention testers' reviews of requirements as a good practice that enhances both the communication and the quality of the requirements, thereby resulting in better alignment of the testing effort. Furthermore, this practice was described as enabling early identification of problems with the test specification avoiding (more expensive) problems later on (C2:62). A systems architect from Company F described that close collaboration between requirements and testing around quality requirements had resulted in 'one area where we have the best alignment' (F4:101).
- **Defining a requirements review responsible (P1.4)** was mentioned as a practice that ensures that requirement reviews are performed (E2:18, F2:114). In addition, for Company F this role was also mentioned as reviewing the quality of the requirements specification (F2:114) and thereby directly addressing the alignment challenge of low quality of the requirements specification (Ch3).
- **Involving domain experts in the requirements definition (P1.5)** was mentioned as a practice to achieve better synchronisation between the requirements and the system capabilities, and thereby support defining more realistic requirements. The expert 'will know if we understand [the requirement] correctly or not' (D3:38), said a system architect. Similar to the previous RE practices, this practice was also mentioned as supporting alignment by enhancing the quality of the requirements (Ch3) which are the basis for software testing.

- **Documentation of requirement decision rationales (P1.6)**, and not just the current requirement version, was suggested as a practice that might facilitate alignment by interviewees from both of the larger companies in our study, namely E and F. 'Softly communicating how we [requirements roles] were thinking' (E3:90) could enhance the synchronisation between project phases by better supporting hand-over between the different roles (F4:39). In addition, the information could support testers in analysing customer defect reports filed a long time after development was completed, and in identifying potential improvements (E3:90). However, the information needs to be easily available and connected to the relevant requirements and test cases for it to be practically useful to the testers (F1:120).

*4.2.2 Validation Practices*

Practices for validating the system under development and ensuring that it is in-line with customer expectations and that the right product is built (IEEE610) include test case reviews, automatic testing of acceptance test cases, and review of prototypes.

- **Test cases are reviewed against requirements (P2.1)** at company F (F14:62). In their new (agile) development processes, the attributes of ISO9126 (ISO9126) are used as a checklist to ensure that not only functional requirements are addressed by the test cases, but also other quality attributes (F14:76).
- **Acceptance test cases defined by customer (P2.2)**, or by the business unit, is practiced at company B**.** The communication with the customer proxy in terms of acceptance criteria for (previously agreed) user scenarios acts as a 'validation that we [software developers] have interpreted the requirements correctly' (B1:117). This practice in combination with full unit test coverage of the code (B1:117) was experienced to address the challenge of achieving full test coverage of the requirements (Ch4, see Section 4.1.4).
- **Reviewing prototypes (P2.3)** and GUI mock-ups was mentioned as an alignment practice applied at company A. With this practice, the product manager in the role as customer proxy validates that the developed product is in-line with the original product intents (A3:153,163). Company partners that develop tailor-made systems using their components may also be involved in these reviews.
- **Management base launch decisions on test reports (P2.4)** was mentioned as an important improvement in the agile way of working recently introduced at Company F. Actively involving management in project decisions and, specifically in deciding if product quality is sufficient for the intended customers was seen as ensuring and strengthening the coordination between customer and business requirements, and testing; 'Management ... have been moved down and [made to] sit at a level where they see what really happens' (F15:89).
- **User/customer testing (P2.5)** is a practice emphasised by company B that apply agile development practices. At regular intervals, executable code is delivered, thus allowing the customer to test and validate the product and its progress (B3: 32, B3:99). This practice is also applied at company E, but with an organisational unit functioning as the user proxy (E3:22). For this practice to be effective the customer testing needs to be performed early on. This is illustrated by an example from company F, namely 'before the product is launched the customer gets to test it more thoroughly. And they submit a lot of feedback. Most are defects, but there are a number of changes coming out of that. That's very late in the process ... a few weeks […] before the product is supposed to be launched' (F1:12). If the feedback came earlier, it could be addressed, but not at this late stage.

*4.2.3 Verification Practices*

Verification ensures that a developed system is built according to the specifications (IEEE610). Practices to verify that system properties are aligned to system requirements include starting verification early to allow time for feedback and change, using independent test teams, re-use of customer feedback obtained from previous projects, and training testers at outsourced or off-shored locations.

- **Early verification (P3.1)** is put forward as an important practice especially when specialised hardware development is involved, as for an embedded product. Verification is then initially performed on prototype hardware (F15:114). Since quality requirements mostly relate to complete system characteristics, early verification of these requirements is harder, but also more important. Company E states: 'If we have performance issues or latency issues or database issues then we usually end up in weeks of debugging and checking and tuning.' (E3:28)
- **Independent test teams (P3.2)** are considered a good practice to reduce bias in interpreting requirements by ensuring that testers are not influenced by the developers' interpretation of requirements. However, this practice also increases the risk of mis-alignment when the requirements are insufficiently communicated since there is a narrower communication channel for requirements-related information. This practice was emphasised especially for companies with safety requirements in the transportation domain (companies C and D); 'due to the fact that this is a fail-safe system, we need to have independency between testers and designers and implementers' (C3:24, similar in C2:39, D2:80), 'otherwise they [test team] might be misled by the development team' (D1:41). Similarly, company F emphasises alternative perspectives taken by an independent team. As a software developer said: 'You must get another point of view of the software from someone who does not know the technical things about the in-depth of the code, and try to get an overview of how it works.' (F13:32)
- **Testers re-use customer feedback from previous projects (P3.3)** when planning the verification effort for later projects (F14:94), thereby increasing the test coverage. In addition to having knowledge of the market through customer feedback, verification organisations often analyse and test competitor products. With a stronger connection and coordination between the verification and business/requirements units, this information could be utilised in defining more accurate roadmaps and product plans.
- **Training off-shore/outsourced testers (P3.4)** in the company's work practices and tools increases the competence and motivation of the outsourced testers in the methods and techniques used by the outsourcing company. This was mentioned by a project manager from Company C as improving the quality of verification activities and the coordination of these activities with requirement (C3:49, 64).

*4.2.4 Change Management Practices*

Practices to manage the (inevitable) changes in software development may mitigate the challenge of maintaining alignment (Ch5, see Section 4.1.5). We identified practices related to the involvement of testing roles in the change management process and also practices connected to product lines as a means to support REVV alignment.
- **Involving testing roles in change management (P4.1)**, in the decision making and in the communication of changes, is a practice mentioned by all companies, but one, as supporting alignment through increased communication and coordination of these changes with the test organisation. '[Testers] had to show their impacts when we [product management] were deleting, adding or changing requirements' (E2:73) and 'any change in requirement ... means involving developer, tester, project manager, requirements engineer; sitting together when the change is agreed, so everybody is aware and should be able to update accordingly' (F8:25). In companies with formalised waterfall processes, a change control board (CCB) is a common practice for making decisions about changes. Company D has weekly meetings of the 'change control board with the customer and we also have more internal change control boards' (D1:106). The transitioning to agile practices affected the change management process at companies E and F. At company F the change control board (CCB) was removed, thus enhancing local control at the expense of control of the whole development chain. As expressed by a process manager in company F: 'I think it will be easy for developers to change it [the requirements] into what they want it to be.' (F12:135) At company E centralised decisions were retained at the CCB (E2:73), resulting in a communication challenge; 'sometimes they [test] don't even know that we [product management] have deleted requirements until they receive them [as deleted from the updated specification]' (E2:73).
- **Product-line requirements practices (P4.2)** are applied in order to reduce the impact of a requirements change. By sharing a common product line (a.k.a. platform), these companies separate between the requirements for the commonality and variability of their products. In order to reduce the impact of larger requirements changes and the risks these entail for current projects, company A 'develop it [the new functionality] separately, and then put that into a platform' (A3:65). Company C use product lines to leverage on invested test effort in many products. When changing the platform version 'we need to do the impact analysis for how things will be affected. And then we do the regression test on a basic functionality to see that no new faults have been introduced.' (C3:55)

*4.2.5 Process Enforcement Practices (P5)*

External requirements and regulations on certain practices affect the motivation and incentive for enforcing processes and practices that support alignment. This is especially clear in company C, which develops safety critical systems. 'Since it is safety-critical systems, we have to show that we have covered all the requirements, that we have tested them.' (C1:6) It is admitted that traceability is costly, but, non-negotiable in their case. 'There is a lot of administration in that, in creating this matrix, but it has to be done. Since it is safety-critical systems, it is a reason for all the documentation involved.' (C1:06) They also have an external assessor to validate that the processes are in place and are adhered to. An alternative enforcement practice was proposed by one interviewee from company F (which does not operate in a safety-critical domain) who suggested that alignment could be achieved by enforcing traceability through integrating process enforcement in the development tools (F14:161) though this had not been applied.

*4.2.6 Tracing Between Artefacts*

The tracing practices between requirements and test artefacts vary over a large range of options from simple mappings between documents to extensive traces between detailed requirements and test cases.

- **Document-level traces (P6.1)** where links are retained between related documents is the simplest tracing practice. This is applied at company A: 'we have some mapping there, between the project test plan and the project requirement specification. But this is a fragile link.' (A2:69)
- **Requirement - test case traces (P6.2)** is the most commonly mentioned tracing practice where individual test cases are traced to individual requirements. This practice influences how test cases are specified: 'It is about keeping the test case a bit less complex and that tends to lead to keep them to single requirements rather than to several requirements.' (F6:123)
- **Using test cases as requirements (P6.3)** where detailed requirements are documented as test cases is another option where the tracing become implicit at the detailed level when requirements and test cases are represented by the same entity. This practice was being introduced at company F. 'At a certain level you write requirements, but then if you go into even more detail, what you are writing is probably very equivalent to a test case.' (F5:113) While this resolves the need for creating and maintaining traces at that level, these test-case requirements need to be aligned to requirements and testing information at higher abstraction levels. 'There will be teams responsible for mapping these test cases with the high-level requirements.' (F10:150) Company A has this practice in place, though not pre-planned but due to test cases being better maintained over time than requirements. 'They know that this test case was created for this requirement some time ago […and] implicitly […] the database of test cases becomes a requirements specification.' (A2:51)
- **Same abstraction levels used for requirements and test specifications (P6.4)** is an alignment practice related to the structure of information. First, the requirements information is structured according to suitable categories. The requirements are then detailed and documented within each category, and the same categorisation used for the test specifications. Company C has 'different levels of requirements specifications and test specifications, top level, sub-system, module level, and down to code' (C3:67), and company D presents similar on the test processes and artefacts (D3:53). It is worth noting that both company C and D develop safety-critical systems. At company F, a project leader described 'the correlation between the different test [levels]' and different requirement levels; at the most detailed level 'test cases that specify how the code should work' and at the next level 'scenario test cases' (F8:16).

*4.2.7 Practice of Traceability Responsible Role (P7)*

For large projects, and for safety-critical projects, the task of creating and maintaining the traces may be assigned to certain roles. In company E, one of the interviewees is responsible for consolidating the information from several projects to the main product level. 'This is what I do, but since the product is so big, the actual checking in the system is done by the technical coordinator for every project.' (E3:54) In one of the companies with safety-critical projects this role also exists; 'a safety engineer […] worked with the verification matrix and put in all the information […] from the sub products tests in the tool and also we can have the verification matrix on our level' (C2:104.)

*4.2.8 Tool Support*

Tool support is a popular topic on which everyone has an opinion when discussing alignment. The tool practices used for requirements and test management vary between companies, as does the tool support for tracing between these artefacts.

- **Tool support for requirements and test management (P8.1)** varies hugely among the companies in this study, as summarised in Table 5. Company A uses a test management tool, while requirements are stored as text. Companies D and E use a requirements management tool for requirements and a test management tool for testing. This was the previous practice at company F too. Company C uses a requirements management tool for both requirements and test, while Company F aims to start using a test management tool for both requirements and testing. Most of the companies use commercial tools, though company A has an in-house tool, which they describe as 'a version handling system for test cases' (A2:208).

Table 5. Tool usage for requirements and test cases, and for tracing between them. For company F the tool set-up prior to the major process change are also given (marked with 'previous').

|  | Requirements tool | Tracing tool | Testing tool |
| --- | --- | --- | --- |
| Requirements | C, D, E, F (previous) |  | F |
| Traces | C | D, E, F (previous) | F |
| Test cases | C |  | A, D, E, F (current and previous) |

- **Tool support for requirements-test case tracing (P8.2)** is vital for supporting traceability between the requirements and test cases stored in the tools used for requirements and test management. Depending on the tool usage, tracing needs to be supported either within a tool, or two tools need to be integrated to allow tracing between them. For some companies, *only manual tracing is supported*. For example, at company D a systems architect describes that it is possible to 'trace requirements between different tools such as [requirements] modules and Word documents' (D3:45). However, a software manager at the same company mentions problems in connecting the different tools and says 'the tools are not connected. It's a manual step, so that's not good, but it works' (D1:111). *Tracing within tools* is practiced at company C where requirements and test cases are both stored in a commercial requirements management tool: 'when we have created all the test cases for a certain release, then we can automatically make this matrix show the links between [system] requirements and test cases' (C1:8). Company F has used the *between-tools practice* 'The requirements are synchronised over to where the test cases are stored.' (F5:19) However, there are issues related to this practice. Many-to-many relationships are difficult to handle with the existing tool support (F2:167). Furthermore, relationships at the same level of detail are easier to handle than across different abstraction levels. One requirements engineer asks for 'a tool that connects everything; your requirement with design documents with test cases with your code maybe even your planning document,' (F5:17). In a large, complex system and its development organisation, there is a need for 'mapping towards all kinds of directions – per function group, per test cases, and from the requirement level' (F11:139).

  Many interviewees had complaints about their tools, and the integration between them. Merely having tool support in place is not sufficient, but it must be efficient and useable. For example, company E have tools for supporting traceability between requirements and test state of connected test cases but 'we don't do it because the tool we have is simply not efficient enough' (E3:57) to handle the test state for the huge amount of verified variants. Similarly, at company E the integration solution (involving a special module for integrating different tools) is no longer in use and they have reverted to manual tracing practices: 'In some way we are doing it, but I think we are doing it manually in Excel sheets' (E2:49).

  Finally, companies moving from waterfall processes towards agile practices tend to find their tool suite too heavy weight for the new situation (E3:89). Users of these tools not only include engineers, but also management, which implies different demands. A test manager states: 'Things are easy to do if you have a lot of hands on experience with the tools but what you really need is something that the [higher level] managers can use' (F10:249).

*4.2.9 Alignment Metrics (P9)*

Measurements can be used to gain control of the alignment between requirements and testing. The most commonly mentioned metrics concern test case coverage of requirements. For example, company C 'measure[s] how many requirements are already covered with test cases and how many are not' (C1:64). These metrics are derived from the combined requirements and test management tool. Companies E and F have a similar approach, although with two different tools. They both point out that, in addition to the metrics, it is a matter of judgment to assess full requirements coverage. 'If you have one requirement, that requirement may need 16 test cases to be fully compliant. But you implement only 14 out of those. And we don't have any system to see that these 2 are missing.' (E3:81) And, 'just because there are 10 test cases, we don't know if [the requirement] is fully covered' (F11:34). Furthermore, there is a versioning issue to be taken into account when assessing the requirements coverage for verification. 'It is hard to say if it [coverage] should be on the latest software [version] before delivery or ...?' (F10:224) The reverse relationship of requirements coverage of all test cases is not always in place or measured. 'Sometimes we have test cases testing functionality not specified in the requirements database.' (F11:133) Other alignment metrics were mentioned, for example, missing links between requirements and tests, number of requirements at different levels (F5:112), test costs for changed requirements (F14:205), and requirements review status (F14:205). Not all of these practices were practiced at the studied companies even though some mentioned that such measures would be useful (F14:219).

### 4.2.10 Job Rotation Practices (P10)

Job rotation was suggested in interviews at companies D and F as a way to improve alignment by extending contact networks and experiences across departments and roles, and thereby supporting spreading and sharing perspectives within an organisation. In general, the interviews revealed that alignment is very dependent on individuals, their experience, competence and their ability to communicate and align with others. The practice of job rotation was mentioned as a proposal for the future and not currently implemented at any of the included companies.

## 4.3 Practices that Address the Challenges

This section provides an overview of the relationships between the alignment challenges and practices identified in this study (and reported in Sections 4.1 and 4.2). The mapping is intended as an initial guide for practitioners in identifying practices to consider in addressing the most pressing alignment challenges in their organisations. The connections have been derived through analysis of the parts of the interview transcripts connected to each challenge and practice, summarised in Table 6 and elaborated next. The mapping clearly shows that there are many-to-many relations between challenges and practices. There is no single practice that solves each challenge. Consequently, the mapping is aimed at a strategic level of improvement processes within a company, rather than a lower level of practical implementation. After having assessed the challenges and practices of REVV alignment within a company, the provided mapping can support strategic decisions concerning which areas to improve. Thereafter, relevant practices can be tailored for use within the specific context. Below we discuss our findings, challenge by challenge.

Table 6. Mapping of practices to the challenges they are found to address. An S represents a suggested, but not implemented practice. Note: a blank cell indicates that no connection was mentioned during the interviews.

| | P1 RE practices | P2 Validation practices | P3 Verification practices | P4 Change management | P5 Process enforcement | P6 Tracing between artefacts | P7 Traceability responsibility role | P8 Tool practices | P9 Alignment metrics | P10 Job rotation |
|---|---|---|---|---|---|---|---|---|---|---|
| Ch1 Aligning goals and perspectives within organisation | P1.1- 1.3, 1.5-1.6 | P2.1, 2.3-2.4 | P3.3 | P4.1 | | P6.1-6.4 | | | | P10 (S) |
| Ch2 Cooperating successfully | P1.2-1.3, 1.5- 1.6 | P2.1, 2.3, 2.4 | P3.1 | P4.1 | | | | | | P10 (S) |
| Ch3 Requirements specification quality | P1.1-1.5 | P2.1, 2.5 | | P4.1 | P5 | P6.2-6.3 | | | P9 | |
| Ch4 VV quality | P1.1- 1.5 | P2.1-2.3, 2.5 | P3.1-3.3 | | P5 | P6.1-6.4 | | | P9 | |
| Ch5 Maintaining alignment when requirements change | | P2.2, P2.5 | | P4.1-4.2 | P5 | P6.1-6.4 | P7 | | P9 | |
| Ch6 Requirements abstraction levels | P1.1, 1.6 | | | | | P6.4 | | | | |

| | | | | | | | | | |
|---|---|---|---|---|---|---|---|---|---|
| Ch7 Traceability | | P2.1 | | | P5 | P6.1-6.4 | P7 | P8.1-8.2 | P9 |
| Ch8 Time and resource availability | | | | P4.1 | P5 | | | | |
| Ch9 Managing a large document space | | | | | | P6.1-6.4 | P7 | P8.1-8.2 | P9 |
| Ch10 Outsourcing of components or testing | P1.1-1.5 | P2.1-2.3 | P3.4 | | | P6.4 | | | |

The practices observed to address the challenge of having **common goals within an organisation (Ch1)** mainly concern increasing the synchronisation and communication between different units and roles. This can be achieved through involving customers and development-near roles in the requirements process (P1.1, P1.2, P1.3, P1.5); documenting requirement decision rationale (P1.6); validating requirements through test case reviews (P2.1) and product managers reviewing prototypes (P2.3); and involving testing roles in change management (P4.1). Goal alignment is also increased by the practice of basing launch decisions made by management on test reports (P2.4) produced by testers. Furthermore, tracing between artefacts (P6.1-6.4) provides a technical basis for supporting efficient communication of requirements. Job rotation (P10) is mentioned as a long-term practice for sharing goals and synchronising perspectives across the organisation. In the mid-term perspective, customer feedback received by testers for previous projects (P3.3) can be re-used as input when defining roadmaps and products plans thereby further coordinating the testers with the requirements engineers responsible for the future requirements.

The challenge of **cooperating successfully (Ch2)** is closely related to the first challenge (Ch1) as being a means to foster common goals. Practices to achieve close cooperation across roles and organisational borders hence include cross-functional involvement (P1.2, P1.5, P2.4) and reviews (P1.3, P2.1, P2.3), feedback through early and continuous test activities (P3.1), as well as, joint decisions about changes in change control boards (P4.1) and documenting requirement decision rationales (P1.6). The former are practices are embraced in agile processes, while the latter practices of change control boards and documentation of rationales were removed for the studied cases when agile processes were introduced. Job rotation (P10), with its general contribution to building networks, is expected to facilitate closer cooperation across organisational units and between roles.

The challenge of achieving good **requirements specification quality (Ch3)** is primarily addressed by the practices for requirements engineering (P1.1-1.5), validation (P2.1, P2.5) and managing requirement changes (P4.1). Some of the traceability practices (P6.2, P6.3) also address the quality of requirements in terms of being well structured and defined at the right level of detail. Furthermore, awareness of the importance of alignment and full requirements coverage may induce and enable organisations in producing better requirements specifications. This awareness can be encouraged with the use of alignment metrics (P9) or enforced (P5) through regulations for safety-critical software and/or by integrating process adherence in development tools.

The challenge of achieving good **validation and verification quality (Ch4)** is addressed by practices to ensure clear and agreed requirements, such as cross-functional reviews (P1.3, P2.1), involving development roles in detailing requirements (P1.2) and customers in defining acceptance criteria (P2.2). Validation is supported by product managers reviewing prototypes (P2.3) and by user/customer testing (P2.5). Verification is improved by early verification activities (P3.1) and through independent testing (P3.2) where testers are not influenced by other engineers' interpretation of the requirements. Complete and up-to-date requirements information is a prerequisite for full test coverage, which can be addressed by requirements engineering practices (P1.1-1.5), testers re-using customer feedback (P3.3) (rather than incorrect requirements specification) and indirectly by traceability practices (P6.1-6.4). The external enforcement (P5) of the full test coverage and alignment metrics (P9) are practices that provide incentives for full test coverage including quality requirements.

**Maintaining alignment when requirements change (Ch5)** is a challenge that clearly connects to change and traceability practices (P4.1-4.2, P6.1-6.4 and P7). However, also the validation practices of having acceptance tests based on user scenarios (P2.2) and user/customer testing (P2.5) address this challenge by providing feedback on incorrectly updated requirements, test cases and/or software. Furthermore, having alignment metrics in place (P9) and external regulations on documentation and traceability (P5) is an incentive to maintain alignment as requirements change.

The challenge of managing **requirements abstraction levels (Ch6)** is addressed by the requirements practice of including the customer in requirements work throughout a project (P1.1) and the tracing practices of matching abstractions levels for requirements and test artefacts (P6.4). Both of these practices exercise the different requirements levels and thereby support identifying mismatches. This challenge is also supported by documentation of requirement decision rationales (P1.6) by providing additional requirements information to the roles at the different abstraction level.

**Traceability (Ch7)** in itself is identified as a challenge in the study, and interviewees identified practices on the information items to be traced (P6.1-6.4), as well as, tools (P8.1-8.2) to enable tracing. In addition, the practice of reviewing test cases against requirements (P2.1) may also support identifying sufficient and/or missing traces. Furthermore, requirements coverage metrics (P9) are proposed as a means to monitor and support traceability. However, as noticed by companies E and F, simple coverage metrics are not sufficient to ensure ample alignment. Process enforcement practices (P5) and assigning specific roles responsible for traceability (P7) are identified as key practices in creating and maintaining traces between artefacts.

The practical aspects of the challenge on **availability of time and resources (Ch8)** are mainly a matter of project management practices, and hence not directly linked to the alignment practices. However, the practice of involving testing roles in the change management process (P4.1) may partly mitigate this challenge by supporting an increased awareness of the verification cost and impact of changes. Furthermore, in companies for which alignment practices are externally enforced (P5) there is an awareness of the importance of alignment of software development, but also an increased willingness to take the cost of alignment including tracing.

The **large document space (Ch9)** is a challenge that can be partly addressed with good tool support (P8.1-8.2) and tracing (P6.1-6.4, P7) practices. The study specifically identifies that a tool that fits a medium-sized project may be very hard to use in a large one. One way of getting a synthesised view of the level of alignment between large sets of information is to characterise it, using quantitative alignment measurements (P9). It does not solve the large-scale problem, but may help assess the current status and direct management attention to problem areas.

**Outsourcing (Ch10)** is a challenge that is related to timing, which is a project management issue, and to communication of the requirements, which are to be developed or tested by an external team. The primary practice to apply to outsourcing is customer communication (P1.1). Frequent and good communication can ensure a common perspective and direction, in particular in the early project phases. In addition, other practices for improved cooperation (P1.2-P1.5, P2.1-P2.3) are even more important when working in different organisational units, times zones, and cultural contexts. Furthermore, in an outsourcing situation the requirements specification is a key channel of communication, often also in contractual form. Thus, having requirements and tests specified at the same level of abstraction (P6.4), feasible for the purpose, is a practice to facilitate the outsourcing. Finally, training the outsourced or off-shored team (P3.4) in company practices and tools also addresses this challenge.

In summary, the interviewees brought forward practices, which address some of the identified challenges in aligning requirements and testing. The practices are no quick-fix solutions, but the mapping should be seen as a guideline to recommend areas for long-term improvement, based on empirical observations of industry practice.

# 5  Discussion

Alignment between requirements and test ranges not only the life-cycle of a software development project, but also company goals and strategy, and affects a variety of issues, from human communication to tools and their usage. Practices differ largely between companies of varying size and maturity, domain and product type, etc. One-size alignment practices clearly do not fit all.

A wide collection of alignment challenges and practices have been identified based on the large amount of experiences represented by our 30 interviewees from six different companies, covering multiple roles, domains and situations. Through analysing this data and deriving results from it, the following general observations have been made by the researchers:

1) the human and organisational sides of software development are at the core of industrial alignment practices
2) the requirements are the frame of reference for the software to be built, and hence the quality of the requirements is critical for alignment with testing activities
3) the large difference in size (factor 20) between the companies, in combination with variations in domain and product type, affects the characteristics of the alignment challenges and applicable practices
4) the incentives for investing in good alignment practices vary between domains

**Organisational and human issues** are related to several of the identified challenges (Ch1, Ch2, Ch8, and Ch10). Successful cooperation and collaboration (Ch2) is a human issue. Having common goals and perspectives for a development project is initially a matter of clear communication of company strategies and goals, and ultimately dependant on human-to-human communication (Ch1). Failures to meet customer requirements and expectations are often related to misunderstanding and misconception; a human failure although technical limitations, tools, equipment, specifications and so on, also play a role. It does not mean that the human factor should be blamed in every case and for each failure. However, this factor should be taken into account when shaping the work conditions for software engineers. These issues become even more pressing when outsourcing testing. Jones et al. (2009) found that failure to align outsourced testing activities with in-house development resulted in wasted effort, mainly due to weak communication of requirements and changes of them**.**

Several of the identified alignment practices involve the human and organisational side of software engineering. Examples include communication practices with customers, cross-role and cross-functional meetings in requirements elicitation and reviews, communication of changes, as well as, a proposed job rotation practice to improve human-to-human communication. This confirms previous research that alignment can be improved by increasing the interaction between testers and requirements engineers. For example, including testers early on and, in particular, when defining the requirements, can lead to improved requirements quality (Uusitalo 2008). However, Uusitalo also found that cross collaboration can be hard to realise due to unavailability of requirements owners and testers on account of other assignments and distributed development (Uusitalo 2008). In general, processes and roles that support and enforce the necessary communication paths may enhance alignment. For example, Paci et al. (2012) report on a process for handling requirements changes through clearly defined communication interfaces. This process relies on roles propagating change information within their area, rather than relying on more general communication and competence (Paci 2012). This also confirms the findings of Uusitalo et al. that increased cross communication reduces the amount of assumptions made by testers on requirements interpretation, and results in an increased reliability of test results and subsequent products (Uusitalo 2008). Similarly, Fogelström et al. (2007) found that involving testers as reviewers through test-case driven inspections of requirements increases the interaction with requirements-related roles and can improve the overall quality of the requirements, thereby supporting alignment. Furthermore, even technical practices, such as tool support for requirements and test management, clearly have a human side concerning degree of usability and usefulness for different groups of stakeholders in an organisation.

**Defining requirements of good quality** (Ch3) is central to enabling good alignment and coordination with other development activities, including validation and verification. This challenge is not primarily related to the style of requirements, whether scenario based, plain textual, or formal. But, rather the quality characteristics of the requirements are important, i.e. being verifiable, clear, complete, at a suitable level of abstraction and up-to-date. This relates to results from an empirical study by Ferguson et al. (2006) that found that unclear requirements have a higher risk of resulting in test failures. A similar reverse relationship is reported by Graham (2002), that clearer and verifiable requirements enable testers to define test cases that match the intended requirements. In addition, Uusitalo et al. (2008) found that poor quality of requirements was a hindrance to maintaining traces from test cases. Sikora et al. (2012) found that requirements reviews is the dominant practice applied to address quality assurance of the requirements for embedded systems and that industry need additional and improved techniques for achieving good requirements quality. Furthermore, requirements quality is related to involving, not only requirements engineers in the requirements engineering, but also VV roles in early stages. This can be achieved by involving non-RE roles in reviews and in detailing requirements. This also contributes to cross-organisational communication and learning, and supports producing requirements that are both useful and used. Uusitalo et al. (2008) found that testers have a different viewpoint that makes them well suited to identifying deficiencies in the requirements including un-verifiability and omissions. Martin et al. (2008) take this approach one step further by suggesting that the requirements themselves be specified as acceptance test cases, which are then used to verify the behaviour of the software. This approach was evaluated through an experiment by Ricca et al. (2009) who found that this helped to clarify and increase the joint understanding of requirements with substantially the same amount of effort. Furthermore, our findings that RE practices play a vital role in supporting REVV alignment confirm previous conclusions that the requirements process is an important enabler for testing activities and that RE improvements can support alignment with testing (Uusitalo 2008).

**Company size** varies largely between the six companies in this study. Similarly, the challenges and practices also vary between the companies. While smaller project groups of 5-10 persons can handle alignment through a combination of informal and formal project meetings. Large-scale projects require more powerful process and tool support to ensure coordination and navigation of the (larger) information space between different phases and hierarchies in the organisation. This was illustrated by different views on the same state-of-the-art requirements management tool. The tool supported alignment well in one medium-sized project (company C), but was frequently mentioned by the interviewees for the largest company (company F) as a huge alignment challenge.

In some cases (e.g. company F), agile practices are introduced to manage large projects by creating several smaller, self-governing and less dependent units. Our study shows that this supports control and alignment at the local level, but, at the expense of global control and alignment (company E). The size-related alignment challenges then re-appear in a new shape, at another level in the organisation. For example, granting development teams mandate to define and change detailed requirements increases speed and efficiency at the team level, but increases the challenge of communicating and coordinating these changes wider within a large organisation.

**The incentives for applying alignment practices**, specifically tracing between requirements and test artefacts, vary across the studied companies. Applying alignment practices seems to be connected to the incentives for enforcing certain practices, such as tracing and extensive documentation. The companies reporting the most rigid and continuously maintained alignment practices are those working in domains where customers or regulatory bodies require such practices. Both of these companies (C and D) have enforced alignment practices in their development including tracing between requirements and tests. Interestingly these are also the companies in our study which apply a traditional and rigorous development model. It is our interpretation that the companies with the most agile, and least rigorous, development processes (A and B) are also the companies which rely heavily on people-based alignment and tracing, rather than on investing in more structured practices. These are also the two companies that do not have tracing between artefacts in place, even partially. While for the remaining companies (E and F) which apply agile-inspired processes, but with structured elements (e.g. eRUP), traceability is in place partly or locally. Our interpretation of the relationship between the included companies concerning incentives and degree of rigour in applying structured alignment practices is illustrated in Figure 4 together with the relative size of their software development. The observed connection between degree of rigour and incentives for alignment are similar to other findings concerning safety-critical development. Namely, that alignment is enabled by more rigorous practices such as concurrently designed processes (Kukkanen 2009) or model-based testing (Nebut 2006, Hasling 2008). Furthermore, companies in safety-critical domains have been found to apply more rigorous processes and testing practices (Runeson 2003). In contrast, neglect of quality requirements, including safety aspects has been found to one of the challenges of agile RE (Cao 2008).

Interestingly, several alignment challenges (e.g. tracing, communicating requirements changes) were experienced also for the companies developing safety-critical software (C and D) despite having tracing in place and applying practices to mitigate alignment challenges (e.g. frequent customer communication, tracing responsible role, change management process involving testers etc.) This might be explained by a greater awareness of the issues at hand, but also that the increased demands posed by the higher levels of quality demands requires additional alignment practice beyond those needed for non-critical software.

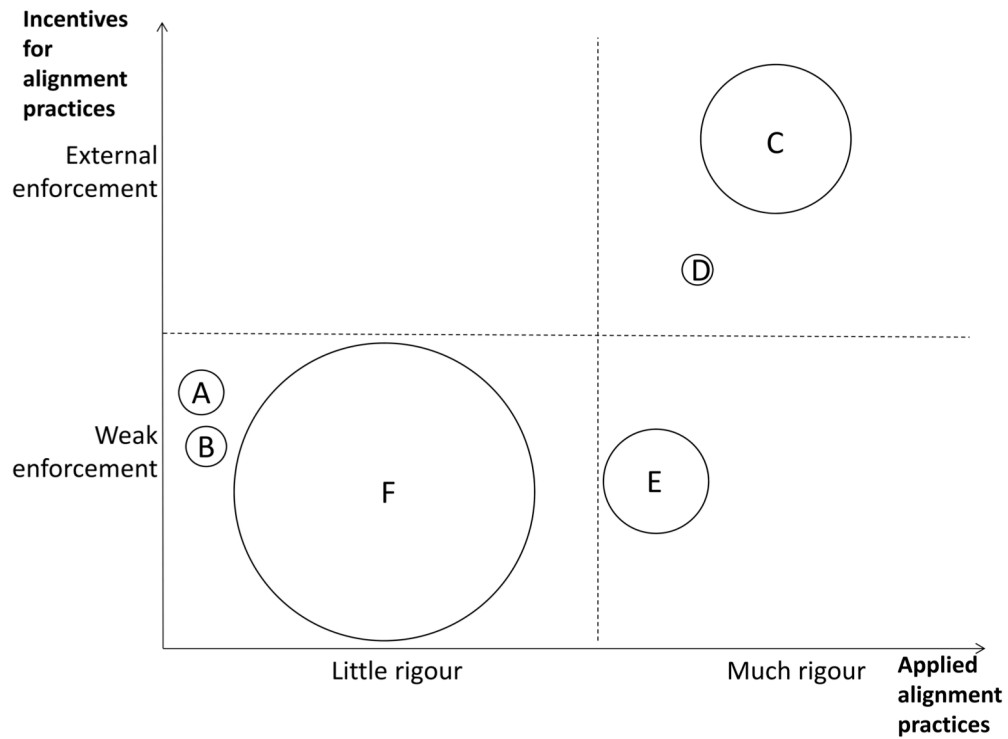

Figure 4. Rough overview of the relationship between the variation factors size, rigour in applying alignment practices and incentive for alignment practices for the studied companies. Number of people in software development is reflected by the relative size of the circle.

When the documentation and tracing practices are directly enforced from outside the organisation, they cannot be negotiated and the cost has to be taken (Watkins 1994). In organisations without these external requirements the business case for investing in these practices needs to be defined, which does not seem to be the case for the studied organisations. Despite the existence of frustration and rework due to bad alignment, the corresponding costs are seldom quantified at any level. Improving alignment involves short term investments in tools, work to recapture traces between large legacies of artefacts, and/or in changed working practices. The returns on these investments are gained mainly in the longer term. This makes it hard to put focus and priority on alignment practices in a short-sighted financial and management culture. Finally, requirements volatility increases the importance and cost to achieve REVV alignment. This need to manage a rate of requirements changes often drives the introduction of agile practices. These practices are strong in team cooperation, but weak in documentation and traceability between artefacts. The companies (C and D) with lower requirements volatility and where development is mainly plan-driven and bespoke, have the most elaborated documentation and traceability practices. In both cases, the practices are enforced by regulatory requirements. However, in our study, it is not possible to distinguish between the effects of different rates of change and the effects of operating in a safety-critical domain with regulations on documentation and traceability.

In summary, challenges and practices for REVV alignment span the whole development life cycle. Alignment involves the human and organisational side of software engineering and requires the requirements to be of good quality. In addition, the incentives for alignment greatly vary between companies of different size and application domain. Future research and practice should consider these variations in identifying suitable practices for REVV alignment, tailored to different domains and organisations.

# 6 Conclusions

Successful and efficient software development, in particular on the large scale, requires coordination of the people, activities and artefacts involved (Kraut 1995, Damian 2005, 2006). This includes alignment of the areas of requirements and test (Damian 2006, Uusitalo 2008, Kukkanen 2009, Sabaliauskaite 2010). Methods and techniques for linking artefacts abound including tracing and use of model-based engineering. However, companies experience challenges in achieving alignment including full traceability. These challenges are faced also by companies with strong incentives for investing in REVV alignment such as for safety critical software where documentation and tracing is regulated. This indicates that the underlying issues lie elsewhere and require aligning of not only the artefacts, but also of other factors. In order to gain a deeper understanding of the industrial challenges and practices for aligning RE with VV, we launched a case study covering six companies of varying size, domain, and history. This paper reports the outcome of that study and provides a description of the industrial state of practice in six companies. We provide categorised lists of (RQ1) industrial alignment challenges and (RQ2) industrial practices for improving alignment, and (RQ3) a mapping between challenges and practices. Our results, based on 30 interviews with different roles in the six companies, add extensive empirical input to the existing scarce knowledge of industrial practice in this field (Uusitalo 2008, Sabaliauskaite 2010). In addition, this paper presents new insights into factors that explain needs and define solutions for overcoming REVV alignment challenges.

We conclude with four high-level observations on the alignment between requirements and testing. Firstly, as in many other branches of software engineering, the *human side* is central, and communication and coordination between people is vital, so also between requirements engineers and testers, as one interviewee said: 'start talking to each other!' (F7:88) Further, the *quality and accuracy of the requirements* is a crucial starting point for testing the produced software in-line with the defined and agreed requirements. Additionally, the *size of the development organisation* and its projects is a key variation factor for both challenges and practices of alignment. Tools and practices may not be scalable, but rather need to be selected and tailored to suit the specific company, size and domain. Finally, alignment practices such as good requirements documentation and tracing seem to be applied for safety-critical development through external enforcement. In contrast, for non-safety critical cases only internal *motivation* exists for the alignment practices even though these companies report facing large challenges caused by misalignment such as incorrectly implemented requirements, delays and wasted effort. For these cases, support for assessing the cost and benefits of REVV alignment could provide a means for organisations to increase the awareness of the importance of alignment and also tailor their processes to a certain level of alignment, suitable and cost effective for their specific situation and domain.

In summary, our study reports on the current practice in several industrial domains. Practical means are provided for recognising challenges and problems in this field and matching them with potential improvement practices. Furthermore, the identified challenges pose a wide assortment of issues for researchers to address in order to improve REVV alignment practice, and ultimately the software engineering practices.

**Acknowledgment**

We want to thank Börje Haugset for acting as interviewer in three of the interviews. We would also like to thank all the participating companies and anonymous interviewees for their contribution to this project. The research was funded by EASE Industrial Excellence Center for Embedded Applications Software Engineering (http://ease.cs.lth.se).